\documentclass[aps,prl,twocolumn,superscriptaddress,showpacs]{revtex4-2}
\usepackage{amsmath,amssymb,graphics,epsfig,epstopdf,color,verbatim,braket,tabularx}
\usepackage[colorlinks,linkcolor=blue,citecolor=blue,urlcolor=blue]{hyperref}
\usepackage{bm}
\usepackage[normalem]{ulem}

\graphicspath{{Figures/}}

\newcommand\COMMENTED[1] {}

\begin{document}

\title{Linear Ensemble-Constrained Quantum Monte Carlo: From Dilute Fermi Gas to Flat-Band Ferromagnetism}

\author{Tu Hong}
\affiliation{Key Laboratory of Artificial Structures and Quantum Control (Ministry of Education)\\
 School of Physics and Astronomy, Shanghai Jiao Tong University, Shanghai 200240, China}

\author{Kun Chen}
\email{chenkun@itp.ac.cn}
\affiliation{Institute of Theoretical Physics, Chinese Academy of Sciences, Beijing, China}

\author{Xiao Yan Xu}
\email{xiaoyanxu@sjtu.edu.cn}
\affiliation{Key Laboratory of Artificial Structures and Quantum Control (Ministry of Education)\\
 School of Physics and Astronomy, Shanghai Jiao Tong University, Shanghai 200240, China}
\affiliation{Hefei National Laboratory, Hefei 230088, China
}

\date{October 18, 2026}

\begin{abstract}
We present a finite‑temperature canonical‑ensemble determinant quantum Monte Carlo algorithm that enforces an exact fermion number and enables numerically stable simulations of correlated lattice fermions. We propose a stabilized QR update that reduces the computational complexity from standard cubic scaling $O(\beta N^3)$ to linear scaling $O(\beta N N_e^2)$ with respect to the system size $N$, where $N_e$ is the particle number. This yields a dramatic speedup in dilute regimes ($N_e \ll N$), opening unbiased access to large-scale simulations of strongly correlated low-density phases. We validate the method on the dilute Fermi gas with onsite Hubbard interactions, observing the suppression of the fermion sign problem in the dilute limit. Furthermore, we apply this approach to a one-dimensional flat-band system, where the canonical ensemble allows for precise control over filling. We reveal a ferromagnetic instability at low temperatures in the half-filling regime. Our linear-scaling approach provides a powerful tool for investigating emergent phenomena in dilute quantum matter.
\end{abstract}

\maketitle

{\it Introduction}\,---\,
Low carrier density regimes of correlated electron systems host a variety of quantum phenomena driven by a common mechanism: the quenching of kinetic energy relative to Coulomb interactions. As the average interparticle spacing---captured by the dimensionless Wigner–Seitz radius $r_s$---increases, the collapsing Fermi energy ($E_F \propto n^{2/d}$) drives the system away from a simple Fermi liquid.
In the clean continuum limit, this evolution produces enhanced effective masses, diverging spin susceptibilities, and ultimately Wigner crystallization at large $r_s$~\cite{Wigner1934,Ceperley1980,Tanatar1989,Drummond2009,Attaccalite2002}.
In complex material environments, such as dilute doped oxides and semiconductor heterostructures, the depressed Fermi energy often becomes comparable to phonon or dielectric modes. This breakdown of the adiabatic approximation facilitates unconventional superconducting phases that defy the standard BCS description, exhibiting nonmonotonic critical temperature dependence and strong-coupling features~\cite{Schooley1964,Lin2014,reyren2007superconducting}.
Furthermore, in lattice systems where low filling coincides with narrow bandwidths or nontrivial topology---including Landau levels and moir\'e superlattices---the kinetic quenching is amplified by band geometry. This interplay can stabilize exotic phases ranging from fractional Chern insulators~\cite{Tang2011,Neupert2011,Sun2011,Haldane2011,Regnault2011,sheng2011fractional} to anomalous Hall crystals, where spontaneous symmetry breaking coexists with a topological phase~\cite{Dong2024,Soejima2024,dong2024theory,dong2403stability,patri2024extended}.
Collectively, these studies emphasize that interaction-driven ordering tendencies, whether in the continuum or on a lattice, are dramatically strengthened in the dilute limit, necessitating nonperturbative approaches to capture the competition between long-range correlations and quantum fluctuations.

From a computational standpoint, unbiased finite-temperature determinant quantum Monte Carlo (DQMC), diagrammatic Monte Carlo, and related methods in this regime face the coupled challenges of achieving fine-grained exact particle-number control at very low densities on feasible lattice sizes, mitigating the fermionic sign problem, and managing the steep growth of computational cost with system size and interaction strength~\cite{Blankenbecler1981,Prokofev2008,chen2019combined,assaad2008world,Loh1990,Troyer2005}. Taking finite-temperature DQMC as an example, the computational complexity is $O(\beta N^3)$, where $N$ is the system size, and $\beta$ is the inverse temperature.  
Numerous optimizations, including the delay update~\cite{sunDelayUpdateDeterminant2024}, the submatrix update~\cite{sunBoostingDeterminantQuantum2025}, and self-learning updates~\cite{liuSelflearningMonteCarlo2017,xuSelflearningQuantumMonte2017,panSelflearningMonteCarlo2025}, enhance performance but fail to alter this underlying scaling. While truncation algorithms can reach $O(\beta N N_e^2)$~\cite{heReachingContinuumLimit2019,gilbreth2021reducing} (with particle number $N_e$), they introduce a systematic error that requires careful control. Therefore, developing an algorithm that is both unbiased and achieves subcubic scaling remains a central challenge in computational many-body physics.

Several studies have imposed the canonical ensemble in DQMC~\cite{Lang1993,Ormand1994,Gilbreth2013,Wang2017,Ouyang2021,Shen2023}. In parallel, other efforts have integrated real-space Fock-state sampling into DQMC, primarily to enhance the efficiency of simulating cold-atom ``snapshots''~\cite{humeniukNumericallyExactMimicking2021}. More recently, the Fock-state DQMC framework~\cite{dingSamplingElectronicFock2025a} was proposed, offering a new path by sampling both auxiliary fields and Fock-state configurations, which is highly flexible for imposing ensemble constraints. Crucially, however, all of these approaches remain limited by the inherent cubic complexity.

In this work, we go beyond the conventional cubic scaling by introducing a numerically stable and efficient scheme for a canonical-ensemble constrained Fock-state DQMC~\cite{dingSamplingElectronicFock2025a}, built upon a stabilized QR update algorithm~\cite{golubMatrixComputations2013}. This advance reduces the computational complexity from $O(\beta N^3)$ down to $O(\beta N N_e^2)$ while remaining numerically exact and unbiased. It enables large-scale and unbiased simulations of dilute systems, making direct validation of cold-atom experiments feasible. Furthermore, because this efficient, fixed-particle framework is structurally similar to the projection version of DQMC, it seamlessly inherits the rich ecosystem of optimizations previously developed there, establishing it as a powerful, general, and unbiased approach.

{\it Ensemble-constrained quantum Monte Carlo}\,---\,
We start with the partition function of the Hubbard model in the canonical ensemble using the Fock-state basis $\ket{\eta} = \prod_{i} (c_{i\uparrow}^{\dagger})^{n_{i\uparrow}} (c_{i\downarrow}^{\dagger})^{n_{i\downarrow}} \ket{0}$~\cite{dingSamplingElectronicFock2025a}:
\begin{equation}
  Z = \mathrm{Tr}_{\eta} [e^{-\beta H}] = \sum_{\eta, \bm{s}} \det \left[P^{\dagger}_{\eta} B_{\bm{s}}(\beta, 0) P_{\eta} \right],
\end{equation}
where $P_\eta$ projects onto the occupied sites of configuration $\eta$, and $B_{\bm{s}}(\beta, 0) = \prod^{L_\tau}_{\ell=1} B_{\bm{s}_\ell}$ is the imaginary-time evolution matrix under auxiliary fields $\bm{s}$ from a discrete Hubbard-Stratonovich transformation.
 The procedure for auxiliary-fields updates is similar to the projection version of DQMC, and the traditional fast update algorithm has a computational cost of $O(\beta N N_e^2)$ to update all space-time lattice points, while the newly developed delay update~\cite{sunDelayUpdateDeterminant2024,sunBoostingDeterminantQuantum2025,duAcceleratingGroundstateAuxiliaryfield2025} and submatrix algorithm~\cite{sunBoostingDeterminantQuantum2025} can further enhance the speed by leveraging computer cache.

However, ensemble-constrained quantum Monte Carlo (EC-QMC) introduces an additional Monte Carlo step for sampling Fock states. To implement the importance sampling of the Fock states, we need to calculate the ratio of weights when we propose the update $\eta \rightarrow \eta'$,
\begin{equation}
  r =\frac{\det[P_{\eta'}^{\dagger}B_{\bm{s}}(\beta,0)P_{\eta'}]}{\det[P_{\eta}^{\dagger}B_{\bm{s}}(\beta,0)P_{\eta}]} .
\end{equation}
To update the Fock state, a method proposed in Ref.~\cite{dingSamplingElectronicFock2025a} maintains a matrix which depends on the full matrix elements of $B_{\bm{s}}(\beta, 0)$.
Because the matrix $B_{\bm{s}}(\beta, 0)$ is singular at low temperatures and needs to be calculated in a numerically stable way,
the complexity of updating the Fock state is at least $O(\beta N^3)$, which is larger than that for updating the auxiliary fields. 

{\it  Linear ensemble-constrained quantum Monte Carlo}\,---\,
To overcome the bottleneck in Fock state updates, we propose a stable QR update algorithm that reduces the computational scaling from cubic to linear, thereby establishing the linear ensemble-constrained quantum Monte Carlo (LEC-QMC).

Any Fock state update involves particle removal and addition. While removing a particle is physically trivial---amounting to the deletion of existing data---its implementation is structurally involved, as it necessitates permuting the columns to reorganize the matrix. In contrast, adding a particle is algorithmically nontrivial: it requires calculating new imaginary-time evolution under the auxiliary field $\bm{s}$. Since the propagator $B_{\bm{s}}(\beta, 0)$ becomes singular at low temperatures, a direct calculation is unstable.

The previous algorithm implicitly relies on the full matrix $B_{\bm{s}}(\beta, 0)$~\cite{dingSamplingElectronicFock2025a}, which amounts to precalculating evolution for all $N$ spatial positions regardless of occupancy. To avoid this, we introduce an ``on-the-fly'' strategy that computes evolution data only for the specific particle being added, precisely when it is needed. This approach, realized via a QR update algorithm~\cite{golubMatrixComputations2013}, fundamentally reduces the computational complexity of the Fock state update.

For \textit{particle addition} at site $j$, we append a column $\bm{p}$ to $P_\eta$. Instead of forming the full propagator, we propagate only the new column $\bm{p}$ through the time slices. To ensure stability, we perform orthogonalization against the existing stable basis $Q_\eta$ (from the decomposition $B_{\bm{s}} P_\eta = Q_\eta R_\eta$) at each stabilization interval. At $\tau=\beta$, this yields an updated decomposition:
\begin{equation}
  Q_{\eta'} = [Q_{\eta} | \bm{q}], \quad
  R_{\eta'} = \begin{pmatrix} R_{\eta} & \bm{v} \\ \mathbf{0} & r_{N_e + 1} \end{pmatrix}.
\end{equation}
The weight ratio simplifies to:
\begin{equation}
  r = \left[\bm{p}^{\dagger}\bm{q}-\bm{p}^{\dagger}Q_{\eta}(P_{\eta}^{\dagger}Q_{\eta})^{-1}P_{\eta}^{\dagger}\bm{q}\right]r_{N_e+1}.
  \label{eq:ratio}
\end{equation}
Crucially, $(P_{\eta}^{\dagger}Q_{\eta})^{-1}$ is already available from the Green's function calculation. The cost is dominated by the propagation of one vector, scaling as $O(\beta N N_e)$.
\textit{Particle removal} is implemented via Givens rotations to restore the upper-triangular form of $R$ after deleting a column, also scaling as $O(\beta N N_e)$.

Canonical sampling is achieved via particle-hole swap updates (one removal followed by one addition). A full sweep of $N_e$ particles thus scales as $O(\beta N N_e^2)$. In the dilute limit where $N_e$ is fixed and $N \to \infty$, the scaling is strictly linear in $N$.

\begin{figure}[t]
\includegraphics[width=8.2cm]{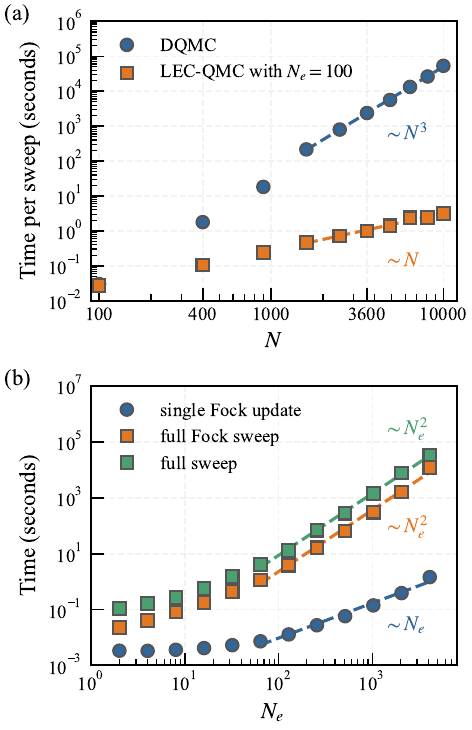}
\caption{\label{fig:time}
Benchmark of the computational scaling of LEC-QMC for the two-dimensional Hubbard model with $\Delta\tau=0.1$, $U/t=2.0$, and $T/t=1.0$.
(a) Average CPU time per full Monte Carlo sweep versus system size $N=L^2$ at fixed particle number $N_e=N_\uparrow+N_\downarrow=100$. The dashed lines are power-law fits, showing the expected $O(N^3)$ scaling for standard DQMC and the linear $O(N)$ scaling for LEC-QMC.
(b) Particle-number scaling at fixed system size $L=100$, with $N_e=2,4,8,\ldots,4096$. The time for a single Fock-state update scales linearly with $N_e$, while both the full Fock-state sweep and the full Monte Carlo sweep scale as $N_e^2$, consistent with the theoretical complexities $O(\beta N N_e)$ and $O(\beta N N_e^2)$, respectively.
}
\end{figure}

{\it Time scaling of the LEC-QMC method}\,---\,
To benchmark the performance of LEC-QMC, we apply it to the repulsive Hubbard model on a two-dimensional square lattice,
\begin{equation}
  H = -t \sum_{\langle i,j \rangle, \sigma} 
  ( c_{i\sigma}^\dagger c_{j\sigma}+\text{H.c.} )
  + U \sum_i n_{i\uparrow} n_{i\downarrow}.
\end{equation}
We first examine the scaling with the system size. Figure~\ref{fig:time}(a) compares the average CPU time per full Monte Carlo sweep between standard DQMC and LEC-QMC at fixed particle number $N_e=N_\uparrow+N_\downarrow=100$, with $U/t=2.0$ and $T/t=1.0$. Standard DQMC exhibits the expected cubic scaling with the system size, $O(N^3)$, whereas LEC-QMC scales linearly with $N$ when $N_e$ and $\beta$ are held fixed. This behavior is consistent with the theoretical sweep complexity $O(\beta N N_e^2)$ and demonstrates the advantage of LEC-QMC in the dilute regime, where $N_e\ll N$.

We further verify the particle-number dependence at fixed system size. As shown in Fig.~\ref{fig:time}(b), for $L=100$ and $N_e=2,4,8,\ldots,4096$, the CPU time for a single Fock-state update scales linearly with $N_e$. Since a full Fock-state sweep contains $O(N_e)$ such updates, its cost scales as $N_e^2$. The full Monte Carlo sweep, including both the Fock-state sweep and the auxiliary-field sweep, follows the same $N_e^2$ scaling. These results confirm the complexity hierarchy: a single Fock-state update costs $O(\beta N N_e)$, while a full sweep costs $O(\beta N N_e^2)$.

The scaling advantage translates into a substantial practical speedup. For the largest dilute systems benchmarked in Fig.~\ref{fig:time}(a), a complete LEC-QMC sweep finishes within seconds under the stated parameters, corresponding to an acceleration of more than 4 orders of magnitude compared with the extrapolated standard-DQMC sweep time. Together, the system-size and particle-number benchmarks establish that LEC-QMC reduces the cubic bottleneck of determinant methods to a particle-number-controlled cost, making large-scale canonical simulations feasible in dilute regimes.

\begin{figure}[t]
\includegraphics[width=8.5cm]{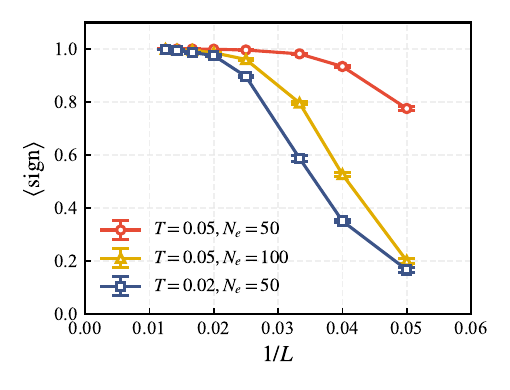}
\caption{\label{fig:sign} Average sign $\langle \text{sign} \rangle$ versus linear system size $L$. On approaching the dilute limit $N = L^2 \gg N_e$, the system asymptotically has no sign problem.}
\end{figure}

{\it Sign problem in the dilute limit}\,---\,
The fermion sign problem is the primary obstacle in QMC. However, physical intuition suggests that in the dilute limit ($n \to 0$), fermions are spatially separated, suppressing the exchange processes responsible for negative signs. We verify this using LEC-QMC. Again, we consider the 2D square lattice Hubbard model with $U/t=2.0$.
Figure~\ref{fig:sign} shows the average sign as a function of linear system size $L$ (where $N=L^2$) for a fixed particle number $N_e$ and fixed temperature. As the system size increases (density decreases), the average sign recovers from near zero to unity. This confirms that the dilute electron gas behaves increasingly like a Boltzmann gas, allowing LEC-QMC to simulate extremely large systems at low densities. 

For the same auxiliary-field decomposition and particle-number
constraints, Fourier-projected canonical DQMC has an average sign
no smaller than that of LEC-QMC. This follows from the triangle
inequality, since Fourier projection sums over the Fock-state
contributions before taking the absolute value of the sampling
weight. Nevertheless, both average signs approach unity in the
dilute limit considered here. The advantage of LEC-QMC in this
regime is its reduced computational cost rather than an improvement
in the average sign.

\begin{figure}[t]
\includegraphics[width=8.5cm]{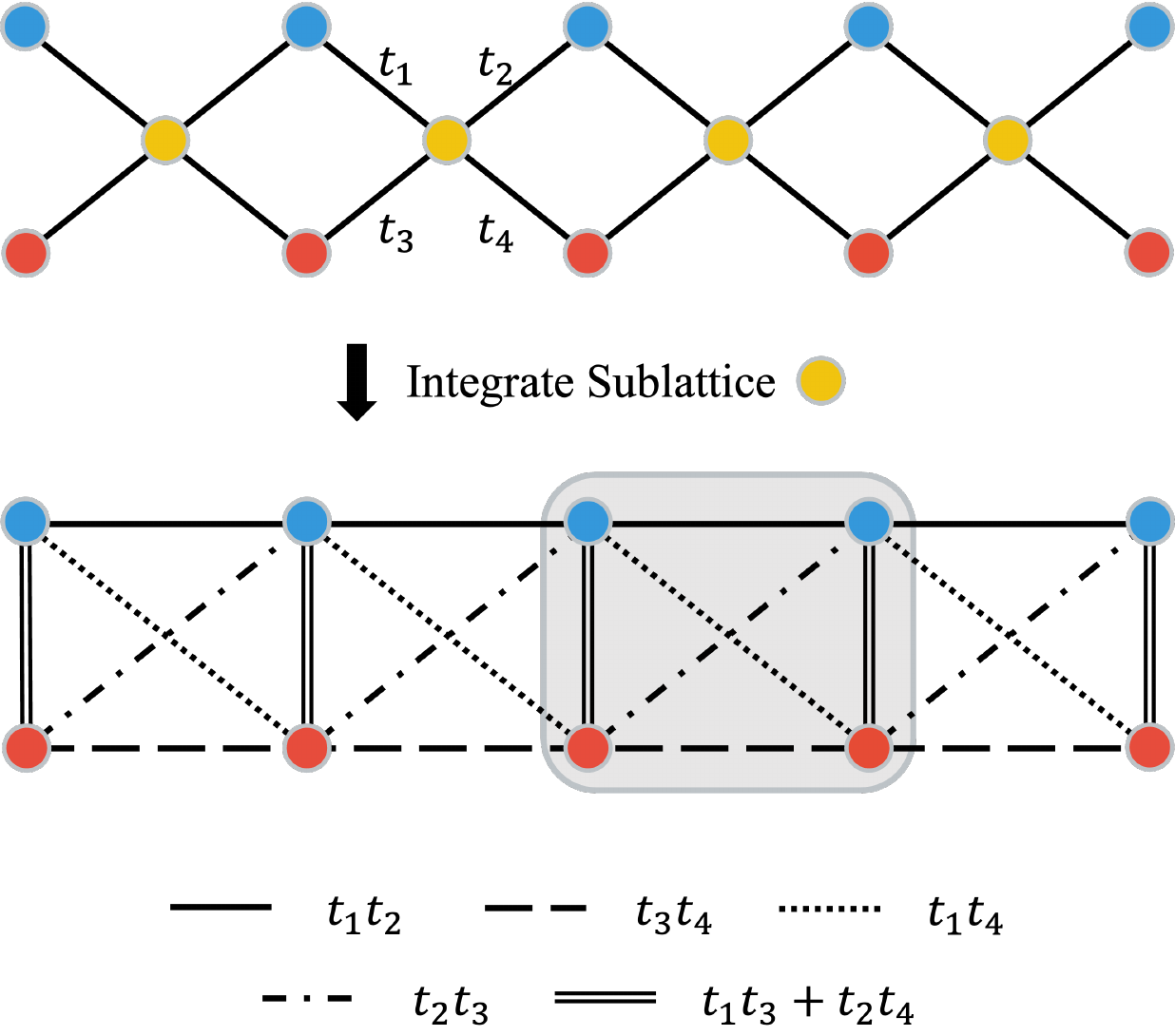}
\caption{\label{fig:flat_band_model} Illustration of the effective flat-band model. The onsite energy of the effective Hamiltonian ($t_1^2 + t_2^2$ for sublattice $A_1$ and $t_3^2 + t_4^2$ for sublattice $A_2$) is omitted for clarity. The gray shaded region indicates the spatial extent of a localized Wannier function, which overlaps with its neighboring Wannier function on two sites.}
\end{figure}

{\it Flat-band ferromagnetism}\,---\,
Flat-band systems represent a singular limit where kinetic energy quenching amplifies interaction effects. To rigorously test our method, we construct a flat-band model using the bipartite crystalline lattice framework in the chiral limit~\cite{cualuguaru2022general}. The original model has three sites in each unit cell, with two of them being the $A$ sublattice, denoted $A_1$ and $A_2$, and the other the $B$ sublattice. By integrating out the $B$ sublattice, we obtain an effective Hamiltonian featuring a perfectly flat lowest band and a gapped upper dispersive band. The illustration of the model is shown in Fig.~\ref{fig:flat_band_model}. In our simulations, we choose the parameters $t_2 = t_3 = 1.0$, $t_1 = t_4 = -0.2$. We consider an onsite Hubbard interaction.
Using the Mielke-Tasaki's theorems on ferromagnetism in the Hubbard model~\cite{mielke1993ferromagnetism}, we can prove that the ground state of our 1D flat-band model at half-filling of the lowest flat band is ferromagnetic. 
However, numerical verification of this phase is notoriously difficult for standard grand-canonical QMC. The infinite compressibility associated with the singular density of states in a flat band makes the particle number extremely sensitive to the chemical potential, rendering precise doping control virtually impossible.

\begin{figure}[t]
\includegraphics[width=8.5cm]{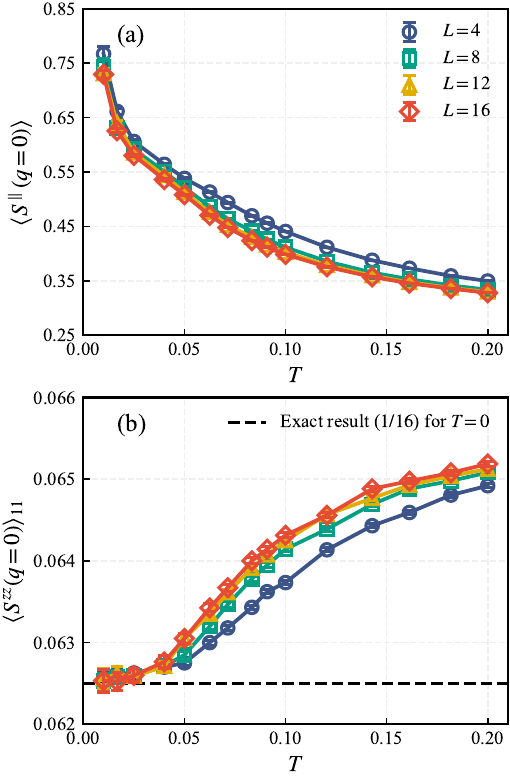}
\caption{\label{fig:ferro} Temperature evolution of (a) in-plane and (b) out-of-plane spin structure factors. The canonical ensemble constraint ($S^z_{\text{tot}} = 0$) confines the magnetic moment to the $XY$ plane, enforcing the sum rule $\langle S^{zz}(q=0)\rangle_{11} = \langle S^{zz}(q=0)\rangle_{22} = - \langle S^{zz}(q=0)\rangle_{12} = -\langle S^{zz}(q=0)\rangle_{21}$. The calculation is done with $t_2 = t_3 = 1.0$, $t_1 = t_4 = -0.2$  and $U/t_2=0.5$. }
\end{figure}

Our LEC-QMC method overcomes this filling-control problem by enforcing the particle number constraint explicitly. We simulate the 1D flat-band model strictly at half-filling, in the sector with total $S^z=0$ ($N_{\uparrow}=N_{\downarrow}$). To diagnose the magnetic order, we compute the equal-time uniform in-plane spin structure factor, defined as $S^{\parallel}(\mathbf{q}=0) = \frac{1}{L} \sum_{i,j} \langle S^x_i S^x_j + S^y_i S^y_j \rangle$.
While the Mermin-Wagner theorem precludes breaking of continuous symmetry at finite temperature in 1D, the ground state ferromagnetism manifests as a divergence of the correlation length as $T \to 0$.
Our results show that $S^{\parallel}(\mathbf{q}=0)$ increases in the low-temperature limit. We also calculate the out-of-plane spin structure factor of the $A_1$ sublattice, and found that it approaches a constant value $1/16$ in the low temperature limit, consistent with the analytical result~\cite{suppl}. This provides unbiased numerical confirmation of the Mielke-Tasaki mechanism and demonstrates the capability of LEC-QMC to resolve ground-state properties in interacting systems with flat bands.

{\it Conclusion}\,---\,
We have developed LEC-QMC, an unbiased quantum Monte Carlo framework that enforces exact particle number conservation while achieving linear computational scaling $O(N)$ in the dilute limit. By combining a canonical-ensemble sampling strategy with a numerically stabilized QR update, we overcome the cubic-scaling bottleneck of conventional determinantal algorithms.
We have demonstrated the power of this approach by explicitly mapping the restoration of sign coherence in the dilute electron gas and by confirming the ferromagnetic ground state of flat-band systems—a regime where standard grand-canonical methods fail due to singular compressibility.

We note that for moderately doped systems, similar to standard DQMC, LEC-QMC still has a severe sign problem in the low-temperature, large-size, and strongly correlated regime. This is especially relevant for flat-band models in higher dimensions, where even a modest linear size already produces a large number of lattice sites, so that away from the dilute limit a severe sign problem persists as in standard DQMC; there, the numerical stability and exact filling control of LEC-QMC remain its key advantages. Whether the milder finite-size effects of higher-dimensional flat bands allow useful finite-temperature results already at small linear sizes is an interesting direction for future work.  
As a separate approximate extension, LEC-QMC can be combined with constrained-path auxiliary-field methods~\cite{zhangConstrainedPathMonte1997, he2019finite} for dense, strongly correlated regimes.
We outline the formalism for applying the constrained path approach within LEC-QMC in the SM~\cite{suppl}.

\begin{acknowledgments}
{\it Acknowledgments}\,---\,
We thank Zhida Song for helpful discussions. We acknowledge the support of the National Natural Science Foundation of China (Grants No. 12447103, No. 12274289, No. 12474245),  the National Key R\&D Program of China (Grants No. 2022YFA1402702, No. 2021YFA1401400, and No. 2024YFA1408604), the Quantum Science and Technology-National Science and Technology Major Project (Grant No. 2021ZD0301902), Yangyang Development Fund, and Shanghai Jiao Tong University 2030 Initiative. The computations in this Letter were run on the Siyuan-1 and $\pi$ 2.0 clusters supported by the Center for High Performance Computing at Shanghai Jiao Tong University.
\end{acknowledgments}

\par
\textit{Data availability}\,---\,
The data supporting the findings of this study are openly available
in Ref.~\cite{Hong2026Data}.

\newpage
\bibliography{LEC-QMC.bib}
\end{document}


\title{Supplementary material for ``Linear Ensemble-Constrained Quantum Monte Carlo: From Dilute Fermi Gas to Flat-Band Ferromagnetism''}

\maketitle

\setcounter{equation}{0}
\setcounter{figure}{0}
\setcounter{table}{0}
\setcounter{page}{1}
\makeatletter
\renewcommand{\thetable}{S\arabic{table}}
\renewcommand{\theequation}{S\arabic{equation}}
\renewcommand{\thefigure}{S\arabic{figure}}
\renewcommand{\bibnumfmt}[1]{[S#1]}
\renewcommand{\citenumfont}[1]{S#1}
\setcounter{secnumdepth}{3}

\section{DETAILS OF LEC-QMC}
In this section, we detail the implementation of the LEC-QMC algorithm, focusing specifically on the efficient update procedures for the Fock state. The core problem we want to solve is obtaining the new QR decomposition $Q_{\eta'} R_{\eta'} = B_{\bm{s}} P_{\eta'}$ efficiently when a particle is added or removed. The main algorithm is based on the QR update algorithm~\cite{golubMatrixComputations2013}.

\begin{figure*}[t]
\includegraphics[width=17cm]{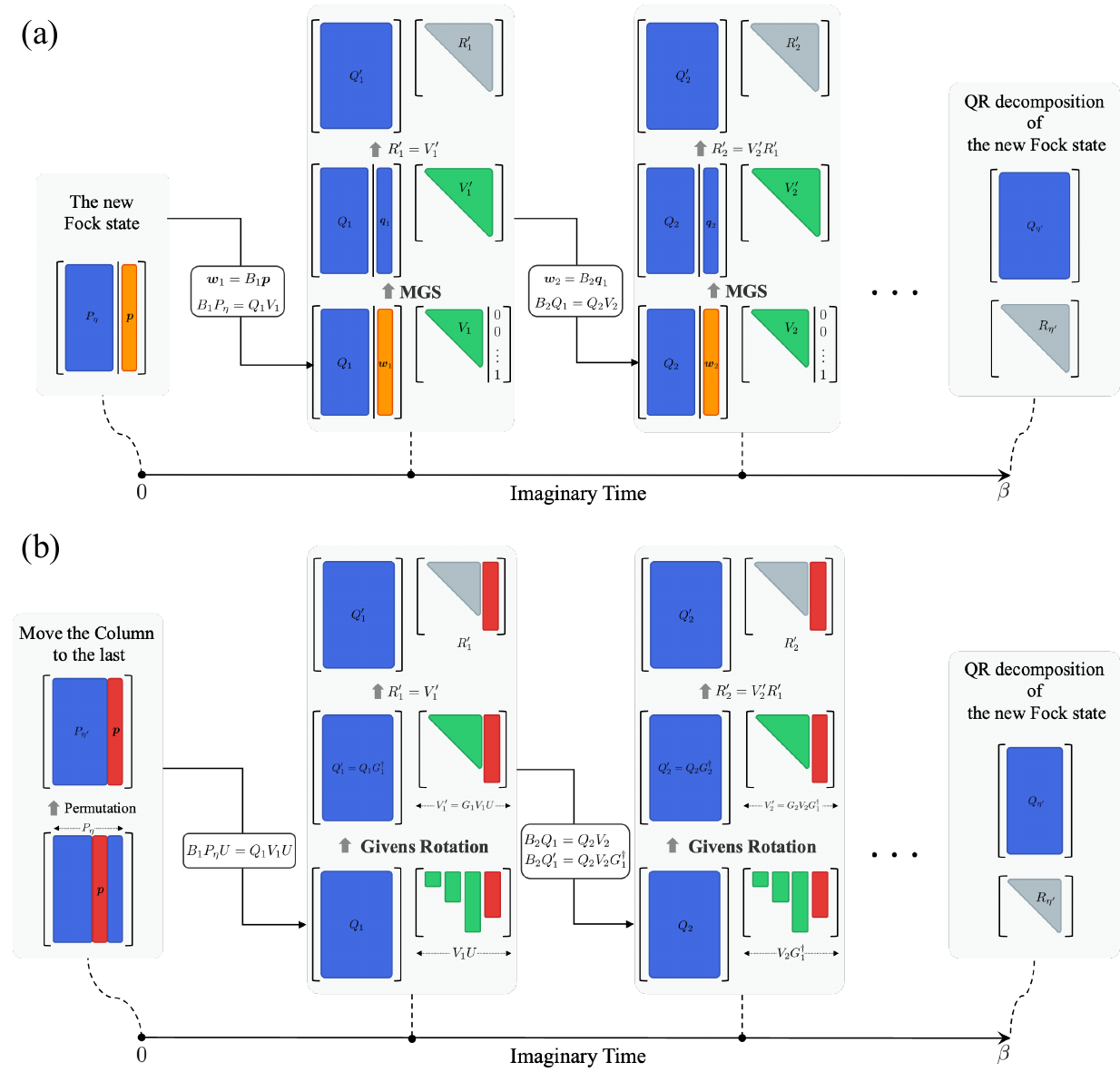}
\caption{\label{fig:fdqmc}The two fundamental update processes in the LEC-QMC method. (a) Adding a new particle (the orange column) to the system. We perform propagation on the new column. At every numerical stabilization step, we perform the modified Gram-Schmidt orthogonalization to get the QR decomposition of the new Fock state. (b) Removing a particle (the red column) from the system. We first apply a permutation to move the column to the last position, then apply Givens rotations at every numerical stabilization step to restore the triangular structure of the matrix. The QR decomposition of the new Fock state is obtained by truncating the last column and row of the matrices.}
\end{figure*}

\subsection{Adding a particle}

Consider the scenario where a new particle is added to the system. This operation mathematically corresponds to appending a new column vector $\bm{p}$ to the existing Fock state matrix $P_{\eta}$:
\begin{equation}
  P_{\eta'} = \left[P_{\eta} \mid \bm{p}\right]
\end{equation}

A straightforward approach to update the decomposition is to propagate the new particle vector $\bm{w} = B_{\bm{s}}(\beta, 0) \bm{p}$ and subsequently perform a modified Gram-Schmidt (MGS) orthogonalization against the existing basis $Q_{\eta}$. This would yield the new basis vector $\bm{q}$ and the updated components:
\begin{gather}
  \bm{v} = Q_{\eta}^{\dagger} \bm{w} \\
  \bm{q} = \frac{\bm{w} - Q_{\eta} \bm{v}}{\|\bm{w} - Q_{\eta} \bm{v}\|} \\
  r_{N_e + 1} = \| \bm{w} - Q_{\eta} \bm{v} \|
\end{gather}
The resulting QR decomposition for the updated state $P_{\eta'}$ would be:
\begin{gather}
  B_{\bm{s}}(\beta, 0) P_{\eta'} = Q_{\eta'} R_{\eta'} \\
  Q_{\eta'} = \left[Q_{\eta} \mid \bm{q}\right], \quad
  R_{\eta'} = \begin{bmatrix}
    R_{\eta} & \bm{v} \\
    \mathbf{0} & r_{N_e + 1}
  \end{bmatrix}
\end{gather}  
Based on this structure, the weight ratio for the particle addition can be derived as:
\begin{align}
  r &=\frac{\det[P_{\eta'}^{\dagger}B_{\bm{s}}(\beta,0)P_{\eta'}]}{\det[P_{\eta}^{\dagger}B_{\bm{s}}(\beta,0)P_{\eta}]} \nonumber\\
    &=\frac{\det[P_{\eta'}^{\dagger}Q_{\eta'}]}{\det[P_{\eta}^{\dagger}Q_{\eta}]}\frac{\det[R_{\eta'}]}{\det[R_{\eta}]} \nonumber\\
    &=\left[\bm{p}^{\dagger}\bm{q}-\bm{p}^{\dagger}Q_{\eta}(P_{\eta}^{\dagger}Q_{\eta})^{-1}P_{\eta}^{\dagger}\bm{q}\right]r_{N_e+1}\nonumber \\
    &=s r_{N_e+1} \label{ratio_htp}
\end{align}
where $s$ denotes the Schur complement in the brackets, and the matrix $(P_{\eta}^{\dagger} Q_{\eta})^{-1}$ is a standard byproduct of the Green's function calculation in projection QMC (PQMC).

While the direct approach above is theoretically sound, the propagator $B_{\bm{s}}(\beta, 0)$ becomes singular at low temperatures, rendering the direct MGS numerically unstable. To overcome this, we perform a layer-by-layer QR update at each stabilization step (we perform numerical stabilization every $\tau_w$ time slices, and in total we perform $n=L_\tau/\tau_w$ stabilization steps), analogous to the standard PQMC sweep. Typically we choose $\tau_w = 10$. The schematic of this process is illustrated in Fig.~\ref{fig:fdqmc}(a), and the detailed procedure is as follows:

\begin{enumerate}
    \item \textbf{First Stabilization Step ($i=1$):}
    \begin{itemize}
        \item \textbf{Propagate and Orthogonalize:}
        Apply the propagator $B_1\equiv B_{\bm{s}}(\tau_w,0)$ to the new particle $\bm{p}$ and orthogonalize the result against the basis $Q_1$:
        \begin{align*}
            \bm{w}_1 &= B_1 \bm{p} \\
            \bm{v}_1 &= Q_1^{\dagger} \bm{w}_1 \\
            \bm{q}_1 &= \frac{\bm{w}_1 - Q_1 \bm{v}_1}{\|\bm{w}_1 - Q_1 \bm{v}_1\|}
        \end{align*}
        \item \textbf{Update Matrices:}
        Construct the augmented basis $Q_1'$ and the ``transfer matrix'' $V_1'$. Note that $R_1'$ is initialized simply as $V_1'$:
        $$
        Q_1' = \left[Q_1 \bigg| \bm{q}_1\right]
        , \quad
        R_1' = V_1' =
        \begin{bmatrix}
            V_1 & \bm{v}_1 \\
            \mathbf{0} & \|\bm{w}_1 - Q_1 \bm{v}_1\|
        \end{bmatrix}
        $$
    \end{itemize}

    \item \textbf{Subsequent Stabilization Steps ($i > 1$):} \\
    Recall that for the original $N_e$ particles, the QR propagation $B_i Q_{i-1} = Q_i V_i$ has already been computed and stored. Here, $B_i \equiv B_{\bm{s}}(i\tau_w,0)$ and $V_i$ serves as the ``transfer matrix'' for the existing basis. We leverage these pre-computed matrices to perform an efficient update:
    
    \begin{itemize}
        \item \textbf{Propagate and Orthogonalize:}
        Propagate the new basis vector $\bm{q}_{i-1}$ from the previous step and orthogonalize it against the current basis $Q_i$:
        \begin{align*}
            \bm{w}_i &= B_i \bm{q}_{i-1}  \\
            \bm{v}_i &= Q_i^{\dagger} \bm{w}_i \quad (\text{Projection onto existing basis}) \\
            \bm{q}_i &= \frac{\bm{w}_i - Q_i \bm{v}_i}{\|\bm{w}_i - Q_i \bm{v}_i\|} \quad (\text{Normalization of residual})
        \end{align*}
        
        \item \textbf{Augment Basis and Transfer Matrix:}
        Construct the updated basis $Q_i'$ and the ``transfer matrix'' $V_i'$ by appending the new components. The block structure preserves the existing dynamics in the upper-left block:
        $$
        Q_i' = \left[Q_i \mid \bm{q}_i\right]
        , \quad
        V_i' =
        \begin{bmatrix}
            V_i & \bm{v}_i \\
            \mathbf{0} & \|\bm{w}_i - Q_i \bm{v}_i\|
        \end{bmatrix}
        $$
        
        \item \textbf{Update Accumulated R Matrix:}
        Update the triangular matrix via the recurrence relation $R_i' = V_i' R_{i-1}'$. Crucially, since the first $N_e$ columns of both $V_i'$ and $R_{i-1}'$ are identical to the pre-computed values, we only need to compute the last column of $R_i'$:
        $$
        [R_i']_{:,N_e+1} = V_i' [R_{i-1}']_{:,N_e+1}
        $$
        This reduces the complexity of this step to $O(N N_e)$.
    \end{itemize}

    \item \textbf{Final Stabilization Step and Update:}
    \begin{itemize}
        \item At the final stabilization step, we obtain the full QR decomposition of the updated Fock state $P_{\eta'}$, with $Q_{\eta'} = Q_n'$ and $R_{\eta'} = R_n'$.
        \item We calculate the weight ratio using Eq.~\eqref{ratio_htp}. 
        \item \textbf{If the update is accepted}, we must update the inverse matrix $(P_{\eta}^{\dagger}Q_{\eta})^{-1}$ to accelerate the next sweep. This is done efficiently using the blockwise inversion formula:
    \end{itemize}
    \begin{widetext}
    \begin{equation}
    (P_{\eta'}^{\dagger}Q_{\eta'})^{-1} =\begin{bmatrix}(P_{\eta}^{\dagger}Q_{\eta})^{-1}+\frac{1}{s}(P_{\eta}^{\dagger}Q_{\eta})^{-1}P_{\eta}^{\dagger}\bm{q}\bm{p}^{\dagger}Q_{\eta}(P_{\eta}^{\dagger}Q_{\eta})^{-1} & -\frac{1}{s}(P_{\eta}^{\dagger}Q_{\eta})^{-1}P_{\eta}^{\dagger}\bm{q}\\
    -\frac{1}{s}\bm{p}^{\dagger}Q_{\eta}(P_{\eta}^{\dagger}Q_{\eta})^{-1} & \frac{1}{s}
    \end{bmatrix}
    \end{equation}
  \end{widetext}
\end{enumerate}

\noindent\textbf{Complexity Analysis:}
The computational cost of a single QR update step is $O(N N_e)$.
 Since the number of numerical stabilization steps ($n$) scales linearly with the inverse temperature, $n \propto \beta$, the total complexity for adding a particle is $O(\beta N N_e)$.

\subsection{Removing a particle}

Removing a particle from the system is structurally more complex than adding one because it disrupts the column ordering. The simplest case is removing the last particle (the last column of $P_{\eta}$), which merely requires deleting the last columns of $Q$ and $R$. 

To remove the $k$-th particle, we employ a strategy of permutation followed by retriangulation. We first apply a permutation matrix $U$ to move the $k$-th column of $P_{\eta}$ to the last position. Since the permutation does not change the weight of the Fock state:
\begin{equation}
    \det\left[ P_{\eta}^\dagger B(\beta, 0) P_{\eta} \right] = \det\left[ U^\dagger P_{\eta}^\dagger B(\beta, 0) P_{\eta} U \right]
\end{equation}
we can proceed by updating the QR decomposition of the permuted system. The process involves applying Givens rotations to restore the upper-triangular structure of the $R$ (or $V$) matrices. The schematic is shown in Fig.~\ref{fig:fdqmc}(b).

\begin{enumerate}
    \item \textbf{First Stabilization Step ($i=1$):}
    \begin{itemize}
        \item \textbf{Permute and Retriangulate:}
        Apply the permutation $U$ to $V_1$ and then a Givens rotation $G_1$ to restore the triangular structure. The updated ``transfer matrix'' $V_1'$ is:
        $$ V_1' = G_1 V_1 U $$
        \item \textbf{Update Basis:}
        The basis is updated to absorb the unitary transformation:
        $$ Q_1' = Q_1 G_1^{\dagger} $$
        \item \textbf{Update R Matrix:}
        For the first stabilization step, the total $R$ matrix is simply:
        $$ R_1' = V_1' $$
    \end{itemize}
    We can verify: $B_1 (P_{\eta} U) = Q_1 (G_1^{\dagger} G_1) V_1 U = Q_1' V_1'$.
    
    \item \textbf{Subsequent Stabilization Steps ($i > 1$):}
    \begin{itemize}
        \item \textbf{Propagate Permutation:}
        The propagator relationship becomes $B_i Q_{i-1}' = Q_i (V_i G_{i-1}^\dagger)$. We need a new Givens rotation $G_i$ to make the term in parentheses upper-triangular.
        \item \textbf{Retriangulate:}
        $$ V_i' = G_i (V_i G_{i-1}^\dagger) $$
        \item \textbf{Update Basis and R Matrix:}
        \begin{align*}
             Q_i' &= Q_i G_i^\dagger \\
             R_i' &= V_i' R_{i-1}' = G_i R_i U
        \end{align*}
    \end{itemize}

    \item \textbf{Final Stabilization Step and Inverse Update:}
    \begin{itemize}
        \item \textbf{Truncation:}
        After the final stabilization step, the particle to be removed is at the last position. The new QR factors $Q_{\eta'}$ and $R_{\eta'}$ are obtained by simply removing the last column of $Q_n'$ and the last row/column of $R_n'$.
        
        \item \textbf{Weight Ratio:}
        The ratio is derived as:
        \begin{align}
            r &= \frac{\det[R_{\eta'}]}{\det[R_{\eta}]} \frac{\det[P_{\eta'}^{\dagger}Q_{\eta'}]}{\det[P_{\eta}^{\dagger}Q_{\eta}]} \nonumber\\
            &= \frac{1}{s r_{N_e}}
        \end{align}
        where $r_{N_e}$ is the last diagonal element of the permuted $R$ matrix.

        \item \textbf{Updating the Inverse Matrix:}
        If the update is accepted, we need to calculate the inverse matrix for the new system $(P_{\eta'}^{\dagger}Q_{\eta'})^{-1}$. 
        First, we compute the inverse of the full permuted matrix, let us denote it as $\mathcal{M}^{-1}$:
        $$
        \mathcal{M}^{-1} = [(P_{\eta} U)^{\dagger} Q_n']^{-1} = G_n (P_{\eta}^{\dagger}Q_{\eta})^{-1} U
        $$
        Considering the block structure of this matrix $\mathcal{M}^{-1}$, we have the following identity relating it to the target inverse $(P_{\eta'}^{\dagger}Q_{\eta'})^{-1}$:
    \end{itemize}
    \begin{widetext}
    \begin{equation}
        \mathcal{M}^{-1} = 
        \begin{bmatrix}
            (P_{\eta'}^{\dagger}Q_{\eta'})^{-1} + \frac{1}{s} \bm{u} \bm{v}^{\dagger} & -\frac{1}{s} \bm{u} \\
            -\frac{1}{s} \bm{v}^{\dagger} & \frac{1}{s}
        \end{bmatrix}
    \end{equation}
    \end{widetext}
    where $\bm{u}$ and $\bm{v}$ are vectors derived from the column updates. The scalar $1/s$ is simply the bottom-right element of $\mathcal{M}^{-1}$.
    Therefore, the inverse matrix for the reduced system can be extracted directly from the sub-blocks of $\mathcal{M}^{-1}$:
    \begin{widetext}
    \begin{equation}
        (P_{\eta'}^{\dagger}Q_{\eta'})^{-1} = \left[\mathcal{M}^{-1}\right]_{1:(N_e-1), 1:(N_e-1)} - \frac{\left[\mathcal{M}^{-1}\right]_{1:(N_e-1), N_e} \times \left[\mathcal{M}^{-1}\right]_{N_e, 1:(N_e-1)}}{\left[\mathcal{M}^{-1}\right]_{N_e, N_e}}
    \end{equation}
    \end{widetext}
\end{enumerate}

The computational cost of a single QR update step for removing a particle is dominated by the Givens rotations with complexity  $O(N N_e)$. Again, since the number of numerical stabilization steps ($n$) scales linearly with the inverse temperature, $n \propto \beta$, the total complexity for removing a particle is also $O(\beta N N_e)$. 

\section{Validation benchmarks}
\subsection{Benchmark against exact diagonalization method}
\begin{figure}[h]
\includegraphics[width=17cm]{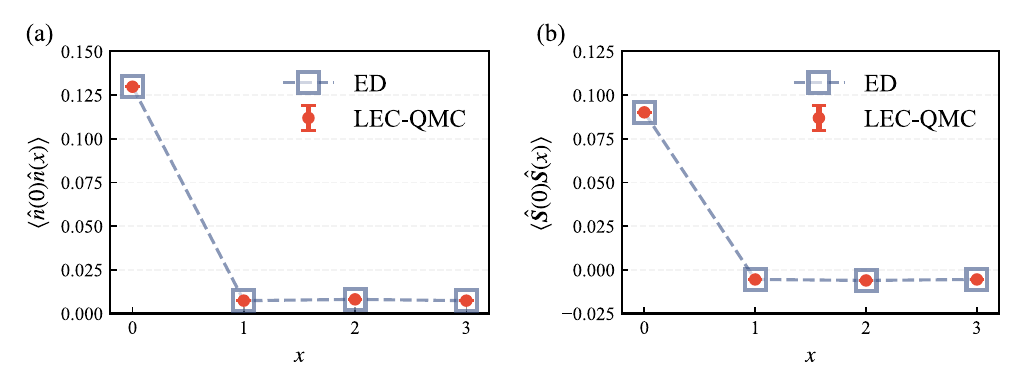}
\caption{\label{fig:ED} Benchmark LEC-QMC results with ED. The parameters used in the simulations are $\Delta \tau = 0.05$, $L = 4$, $N_e = 2$, $T=0.05$ and $U = 2.0$.}
\end{figure}

To ensure the accuracy of LEC-QMC, we perform a benchmark for the 2D repulsive Hubbard model. We choose the system with size $L = 4$ and particle number $N_e = N_{\uparrow} + N_{\downarrow} = 2$, which can be benchmarked by the exact diagonalization (ED) method. We calculated the spin correlation function $\langle \hat{\bm{S}}(\bm{r}_i) \hat{\bm{S}}(\bm{r}_j) \rangle$ and the charge density correlation function $\langle \hat{n}(\bm{r}_i) \hat{n}(\bm{r}_j) \rangle$. 
The results are shown in Fig.~\ref{fig:ED}, which agree well with ED.

\subsection{Benchmark against other canonical ensemble methods}

We further benchmark LEC-QMC against several established ways of imposing or approximating the canonical ensemble. This comparison provides an independent check that the QR-stabilized LEC-QMC formulation samples the correct canonical distribution and does not introduce any systematic bias.

We consider the repulsive Hubbard model on a $6\times 6$ square lattice with $U/t=2$ and fixed particle numbers $N_\uparrow=N_\downarrow=18$. We compare four methods: (i) Fourier projection, which evaluates the canonical trace by projecting from the grand-canonical ensemble~\cite{ormand1994demonstration}, it is unbiased with computational complexity $\mathcal{O}(\beta N^3)$; (ii) penalty projection, where fluctuations of $N_\uparrow$ and $N_\downarrow$ are separately suppressed by adding two corresponding penalty terms to the Hamiltonian~\cite{Wang2017}, for which a large penalty is needed to approach the unbiased results, with computational complexity $\mathcal{O}(\beta N^3)$; (iii) the original Fock-state DQMC (EC-QMC) method~\cite{dingSamplingElectronicFock2025a}, which is numerically unstable in the low temperature region and with computational complexity $\mathcal{O}(\beta N^3)$; and (iv) the present unbiased LEC-QMC method with computational complexity $\mathcal{O}(\beta N N_e^2)$. In this benchmark, the imaginary time Trotter decomposition step is chosen as $\Delta\tau = 0.1$, and the numerical stabilization interval is chosen as $\tau_w =10$. We perform the same number of sweeps (32,000).

\begin{figure}[t]
    \centering
    \includegraphics[width=0.6\linewidth]{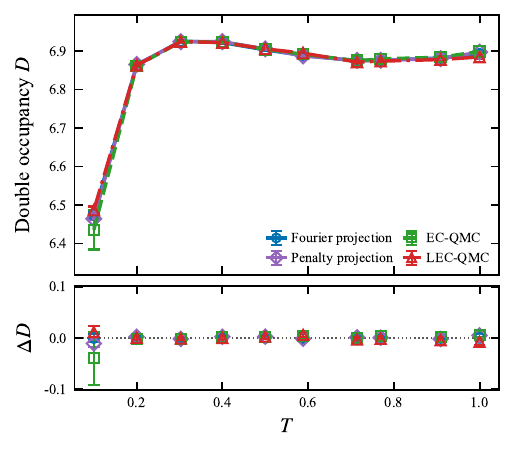}
    \caption{
    Benchmark of the double occupancy in the canonical ensemble for the $6\times 6$ repulsive Hubbard model with $U/t=2$ and $N_\uparrow=N_\downarrow=18$. We compare Fourier projection, penalty projection, EC-QMC, and LEC-QMC. The LEC-QMC results agree with Fourier projection within statistical uncertainty. The lower panel compares deviations of the results from the Fourier projection results, showing that both LEC-QMC and penalty projection overlap with Fourier projection results, while EC-QMC has large errorbars in the low temperature region.
    }
    \label{fig:canonical_double_occupancy}
\end{figure}

Figure~\ref{fig:canonical_double_occupancy} shows the total double occupancy $D=\sum_i \langle n_{i\uparrow} n_{i\downarrow} \rangle$ as a function of temperature. 
All four methods agree with each other in the high temperature region. In the low temperature region, both penalty projection and LEC-QMC agree with the Fourier projection results. However, the EC-QMC results show a small deviation with large error bars in the low temperature region.

\begin{figure}[t]
    \centering
    \includegraphics[width=0.6\linewidth]{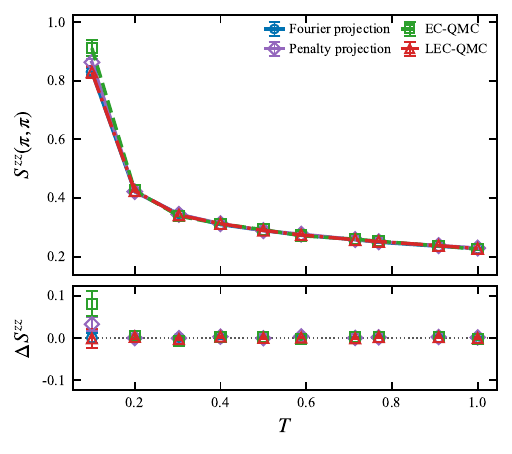}
    \caption{
    Benchmark of the antiferromagnetic spin structure factor $S^{zz}(\pi,\pi)$ for the same system as in Fig.~\ref{fig:canonical_double_occupancy}. The lower panel compares deviations of the results from the Fourier projection results, showing that only LEC-QMC overlaps with Fourier projection results at all temperatures, while penalty projection and EC-QMC deviate from the Fourier projection results in the low temperature region. This comparison highlights the improved numerical stability of the QR-stabilized LEC-QMC formulation for correlation functions.
    }
    \label{fig:canonical_szsz}
\end{figure}

We also compare the antiferromagnetic spin structure factor $S^{zz}(\pi,\pi)$ in Fig.~\ref{fig:canonical_szsz}. The LEC-QMC results remain consistent with the Fourier-projection data at all temperatures. In contrast, both the EC-QMC and penalty projection results show a more visible deviation in this benchmark in the lowest temperature region. Since the original EC-QMC update relies on manipulating the full imaginary-time evolution matrix, the long imaginary-time propagation can become ill-conditioned at low temperatures. The QR-stabilized update in LEC-QMC avoids explicitly forming such an unstable full propagator and instead maintains well-conditioned decompositions throughout the imaginary-time evolution.

These comparisons confirm that LEC-QMC reproduces canonical-ensemble observables obtained from independent projection methods, while providing improved numerical stability over the original Fock-state formulation. Together with the scaling benchmarks, this demonstrates that the present method is both computationally efficient and statistically reliable for fixed-particle-number simulations.

\section{Efficiency benchmarks}

In this section, we provide a set of efficiency benchmarks for LEC-QMC. We first compare the wall-clock time per Monte Carlo sweep with standard DQMC across a broad filling range. We then examine whether the per-sweep acceleration translates into improved statistical efficiency for physical observables by comparing error bars as functions of both the number of sweeps and the total wall-clock time. Throughout this section, we use the two-dimensional repulsive Hubbard model as the benchmark system.

\subsection{Timing comparison with DQMC across filling}

We compare the wall-clock time per Monte Carlo sweep between LEC-QMC and standard DQMC over a broad range of filling fractions. This benchmark clarifies the practical regime in which LEC-QMC provides a computational advantage and also shows the crossover behavior in the dense regime.

We consider square lattices with linear sizes $L=10,20,30,50$, interaction strength $U/t=2$, inverse temperature $\beta t=1$, and Trotter step $\Delta\tau=0.1$. For each fixed system size, we vary the particle number and measure the average CPU time per sweep for both standard DQMC and LEC-QMC. The filling is defined as $n=N_e/N$, where $N=L^2$ is the number of lattice sites and $N_e=N_\uparrow+N_\downarrow$ is the total particle number. We focus on the spin-balanced case, $N_\uparrow=N_\downarrow$; with this convention, half filling corresponds to $n=1$.

\begin{figure}[t]
    \centering
    \includegraphics[width=0.95\linewidth]{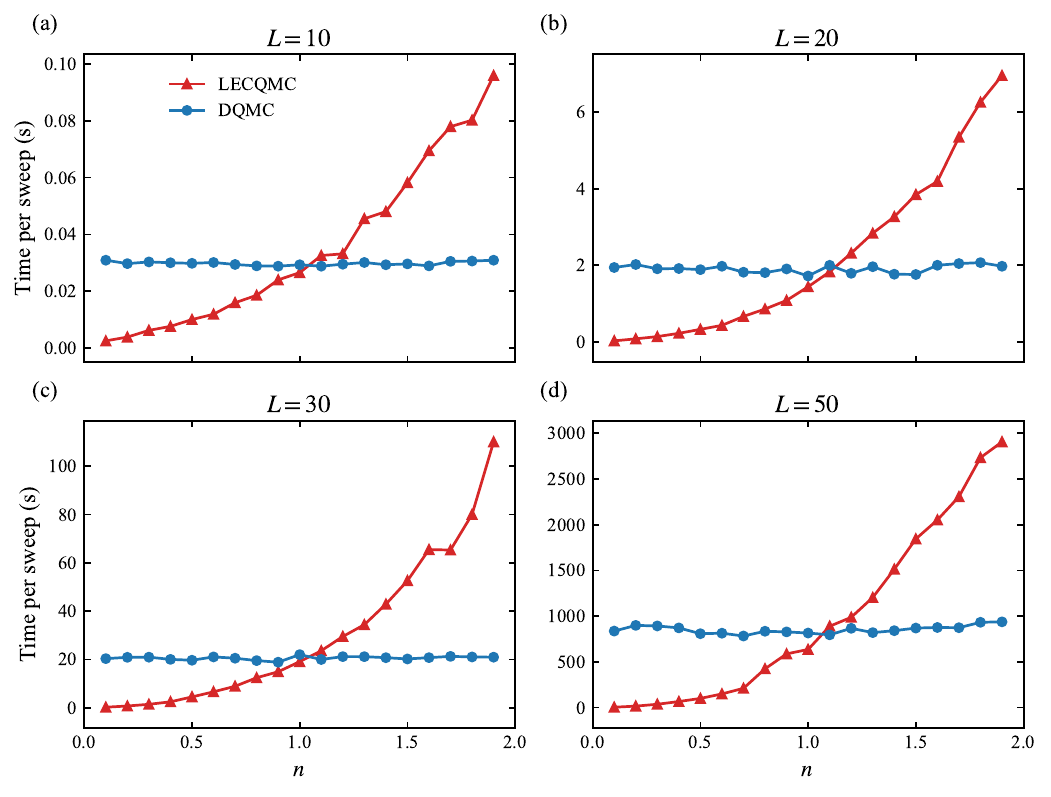}
    \caption{
    Timing comparison between LEC-QMC and standard DQMC across filling fractions for the two-dimensional Hubbard model. The parameters are $U/t=2$, $\beta t=1$, and $\Delta\tau=0.1$. Panels correspond to different system sizes $L=10,20,30,50$. Standard DQMC shows only weak filling dependence, while the LEC-QMC sweep time in the electron representation increases with particle number, consistent with the expected $O(\beta N N_e^2)$ scaling. LEC-QMC is faster in the dilute and below-half-filled regimes. The crossover above half filling reflects the use of the electron representation in this benchmark; the high-filling regime can alternatively be formulated as a dilute-hole problem.
    }
    \label{fig:time_filling_dqmc_lecqmc}
\end{figure}

The results are shown in Fig.~\ref{fig:time_filling_dqmc_lecqmc}. For standard DQMC, the sweep time depends only weakly on filling, as expected from its dominant $O(\beta N^3)$ matrix operations. By contrast, the LEC-QMC sweep time increases with particle number, consistent with the $O(\beta N N_e^2)$ full-sweep scaling discussed in the main text and Sec.~I. In the dilute regime and throughout the range below half filling, LEC-QMC is faster than standard DQMC. The advantage is most pronounced at low density, where the active particle number is much smaller than the system size.

As the filling is increased toward and beyond half filling, the speedup gradually decreases and the two methods eventually cross in the present implementation. This occurs because, in the electron representation used for Fig.~\ref{fig:time_filling_dqmc_lecqmc}, the active matrix dimension grows with the electron number. When $N_e=O(N)$, the asymptotic scaling advantage of LEC-QMC is reduced and the cost becomes comparable to $O(\beta N^3)$. In addition, the auxiliary-field update in LEC-QMC follows a PQMC-like formulation and maintains several rectangular matrices associated with the Fock-state sector, which introduces a larger constant prefactor than standard DQMC.

The high-filling side of this benchmark can be treated more efficiently by switching from an electron representation to a hole representation. This change of variables does not require particle-hole symmetry: for a general lattice Hamiltonian, the particle-hole transformation maps the original electron problem above half filling to an equivalent hole problem below half filling, generally with transformed parameters and possible additional one-body or constant terms. The important algorithmic point is that the number of active particles sampled by LEC-QMC is then the number of holes rather than the number of electrons. Therefore, the crossover shown in Fig.~\ref{fig:time_filling_dqmc_lecqmc} is representation-dependent rather than a fundamental limitation of LEC-QMC. By using the electron representation below half filling and the hole representation above half filling, the relevant active particle number is reduced to $\min(N_e,2N-N_e)$, allowing LEC-QMC to retain its scaling advantage away from the immediate vicinity of half filling.

\subsection{Comparison among DQMC, EC-QMC, and LEC-QMC}

We next compare three methods: standard DQMC, the original Fock-state DQMC formulation, which we denote as EC-QMC here~\cite{dingSamplingElectronicFock2025a}, and the present LEC-QMC method. Both EC-QMC and LEC-QMC sample auxiliary-field configurations together with Fock-state configurations and therefore impose the particle-number constraint directly in the sampled ensemble. The key difference is that EC-QMC relies on the full imaginary-time evolution matrix for Fock-state updates, while LEC-QMC replaces this step by the QR-stabilized update described in Sec.~I.

\begin{figure}[t]
    \centering
    \includegraphics[width=0.95\linewidth]{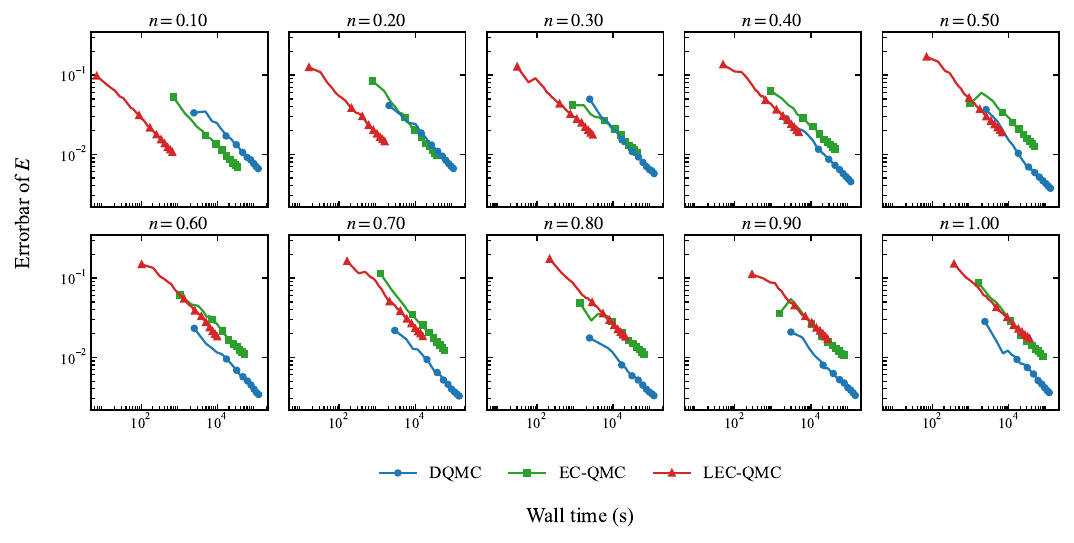}
    \caption{
    Statistical error bar of the total energy as a function of wall-clock time for DQMC, EC-QMC, and LEC-QMC on a log-log scale. The benchmark is performed for the two-dimensional repulsive Hubbard model on an $L=20$ square lattice with $U/t=2$ and $\beta t=1$, for fillings $n=0.1,0.2,\cdots,1.0$. LEC-QMC is more efficient than EC-QMC, or comparable to it at high filling, showing that the QR-stabilized formulation improves the original Fock-state update without increasing the statistical cost.
    }
    \label{fig:fdqmc_error_time}
\end{figure}

Figure~\ref{fig:fdqmc_error_time} compares the statistical error bar of the total energy as a function of wall-clock time for the two-dimensional repulsive Hubbard model on an $L=20$ square lattice with $U/t=2$ and $\beta t=1$. We vary the total filling from $n=0.1$ to $n=1.0$.

Across the full filling range shown in Fig.~\ref{fig:fdqmc_error_time}, LEC-QMC is generally more efficient than EC-QMC. The improvement is most visible at low and intermediate fillings, where the QR-stabilized update reduces the cost of Fock-state sampling. At high filling, the performance of the two Fock-state methods becomes closer because the active particle number is no longer small.

This comparison allows us to focus below on the more stringent benchmark between LEC-QMC and standard DQMC. Standard DQMC does not explicitly sample Fock-state configurations and is highly optimized for auxiliary-field sampling, so it provides a natural reference for assessing the practical efficiency of LEC-QMC. We also note that standard DQMC works in the grand canonical ensemble, whereas LEC-QMC works in a constrained ensemble such as the canonical ensemble. To simulate a fixed-particle system with standard DQMC, one must fine-tune the chemical potential, whereas LEC-QMC has no such requirement. In this benchmark, we neglect this fine-tuning cost for standard DQMC.

\subsection{Energy error bars: LEC-QMC versus DQMC}

We now compare LEC-QMC directly with standard DQMC for the total energy. The simulations are performed for the two-dimensional repulsive Hubbard model with $U/t=2$ and $\beta t=1$. We consider system sizes $L=10,20,30,40$ and total fillings $n=0.2,0.3,\cdots,0.7$. All simulations are spin balanced, $N_\uparrow=N_\downarrow$.

\begin{figure}[t]
    \centering
    \includegraphics[width=0.95\linewidth]{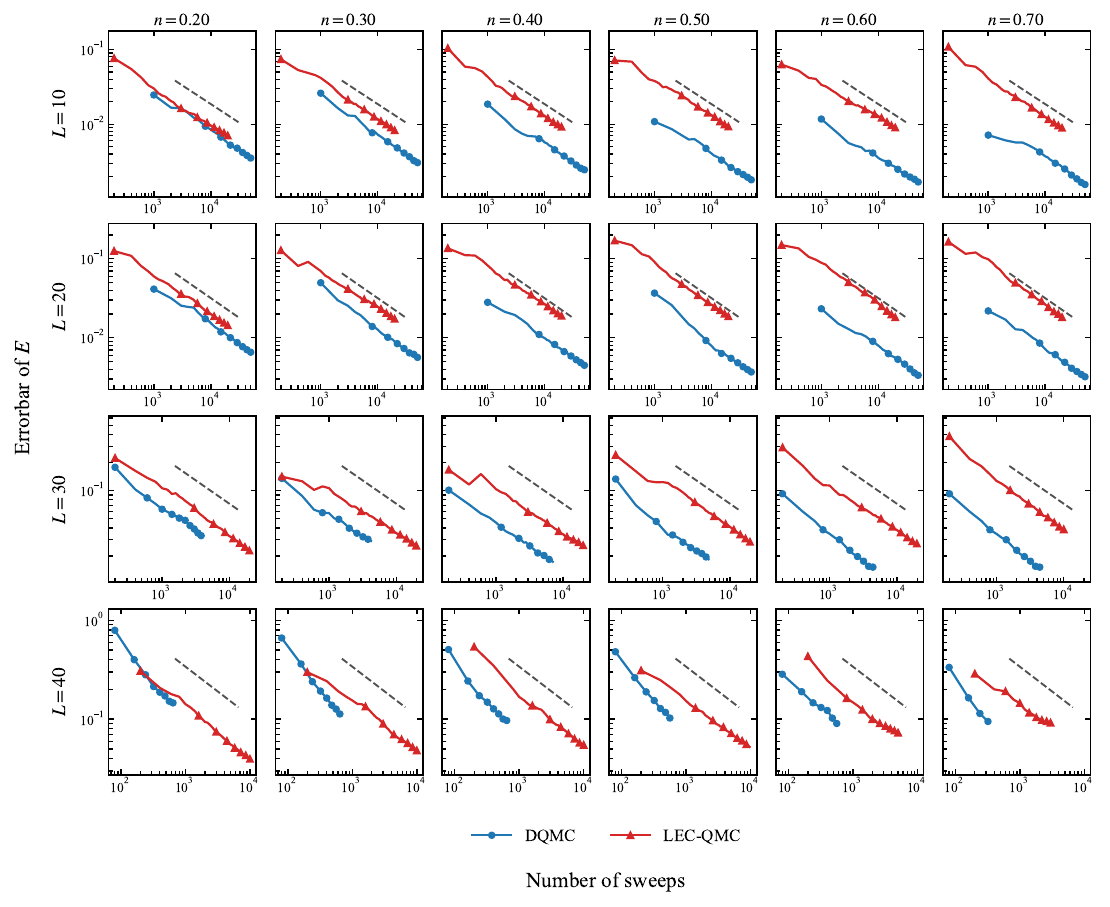}
    \caption{
    Statistical error bar of the total energy as a function of the number of Monte Carlo sweeps for standard DQMC and LEC-QMC on a log-log scale. The model is the two-dimensional repulsive Hubbard model with $U/t=2$ and $\beta t=1$. We compare system sizes $L=10,20,30,40$ and fillings $n=0.2,0.3,\cdots,0.7$. The gray dashed line indicates the ideal $N_{\rm sweep}^{-1/2}$ scaling expected for statistically independent samples. The error bars of both methods decrease approximately with this scaling, indicating that the additional Fock-state sampling in LEC-QMC does not lead to an anomalously slow convergence with sweep number. At fixed sweep number, DQMC often has smaller error bars, especially at larger filling.
    }
    \label{fig:energy_error_sweeps}
\end{figure}

Fig.~\ref{fig:energy_error_sweeps} shows the statistical error bar of the total energy as a function of the number of Monte Carlo sweeps. At fixed sweep number, standard DQMC often has smaller error bars, especially at larger filling. This behavior is expected because LEC-QMC samples not only auxiliary-field configurations but also Fock-state configurations, and this additional sampling sector can increase the variance.

\begin{figure}[t]
    \centering
    \includegraphics[width=0.95\linewidth]{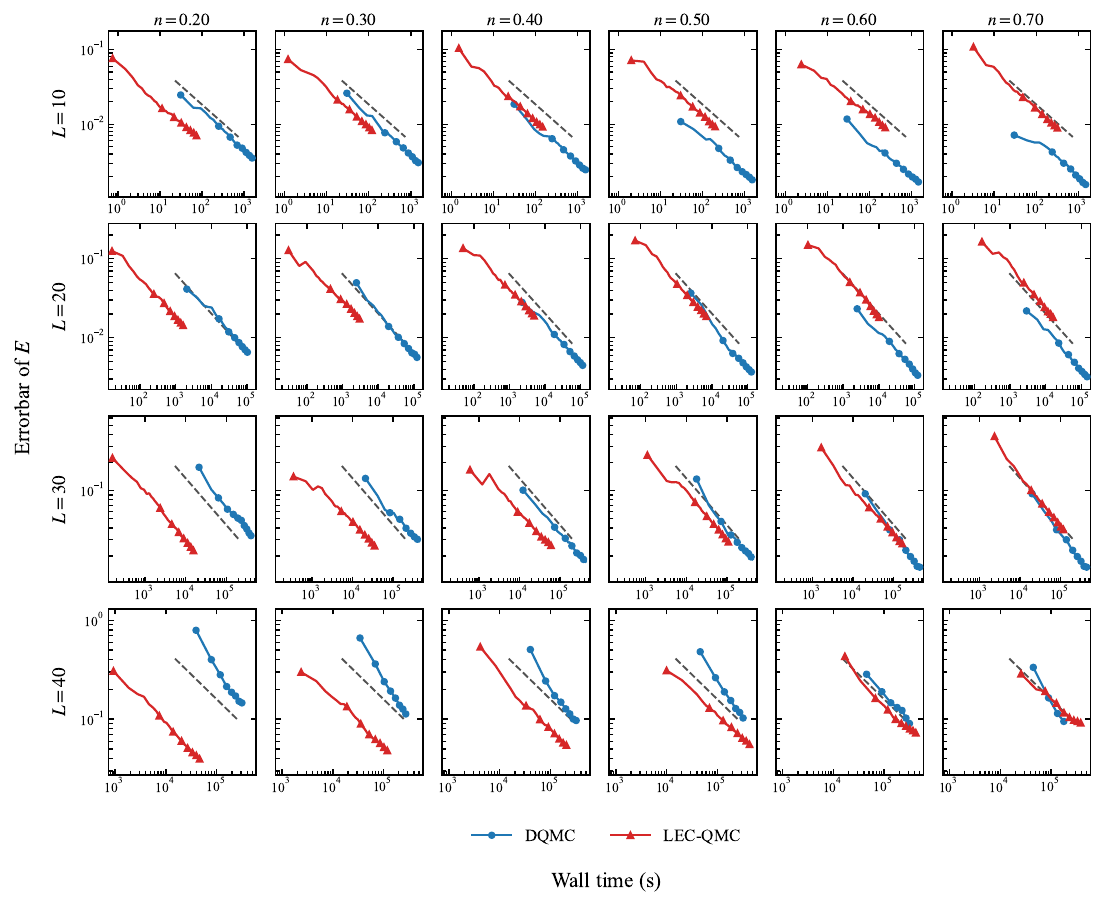}
    \caption{
    Statistical error bar of the total energy as a function of wall-clock time for the same parameters as in Fig.~\ref{fig:energy_error_sweeps}, shown on a log-log scale. The gray dashed line indicates the ideal $t_{\rm wall}^{-1/2}$ scaling expected when the effective number of independent samples grows linearly with wall-clock time. After accounting for the computational cost per sweep, LEC-QMC shows a clear advantage in the dilute and large-system regimes. The filling at which the two methods become comparable shifts to higher density as the system size increases, reflecting the increasing importance of the system-size scaling advantage of LEC-QMC.
    }
    \label{fig:energy_error_time}
\end{figure}

The practical comparison should also include the computational cost of each sweep. Fig.~\ref{fig:energy_error_time} shows the same energy error bars plotted against the total wall-clock time. In this representation, the advantage of LEC-QMC becomes clear in the dilute and large-system regimes. Although DQMC may converge somewhat faster per sweep, each DQMC sweep is much more expensive for large $N$. As a result, LEC-QMC reaches the same statistical accuracy in significantly less wall-clock time at low density.

The density range where LEC-QMC is advantageous expands with system size. From Fig.~\ref{fig:energy_error_time}, the two methods become comparable around $n\simeq 0.4$ for $L=10$, around $n\simeq 0.5$ for $L=20$, around $n\simeq 0.6$ for $L=30$, and around $n\simeq 0.7$ for $L=40$. This trend is consistent with the scaling advantage of LEC-QMC: as the lattice size grows, the reduction from cubic scaling in the full system size to particle-number-controlled scaling becomes increasingly important.

These results demonstrate that the per-sweep timing advantage of LEC-QMC is not offset by a loss of statistical efficiency. Instead, in the low-density and large-system regimes targeted by the present algorithm, LEC-QMC provides a genuine wall-clock speedup for obtaining statistically converged observables.

\subsection{Spin-structure-factor error bars: LEC-QMC versus DQMC}

We also compare the statistical efficiency for the antiferromagnetic spin structure factor $S^{zz}(\pi,\pi)$. This observable provides a complementary benchmark because spin correlations are generally more sensitive to long-distance fluctuations and sampling quality than the total energy.

\begin{figure}[t]
    \centering
    \includegraphics[width=0.95\linewidth]{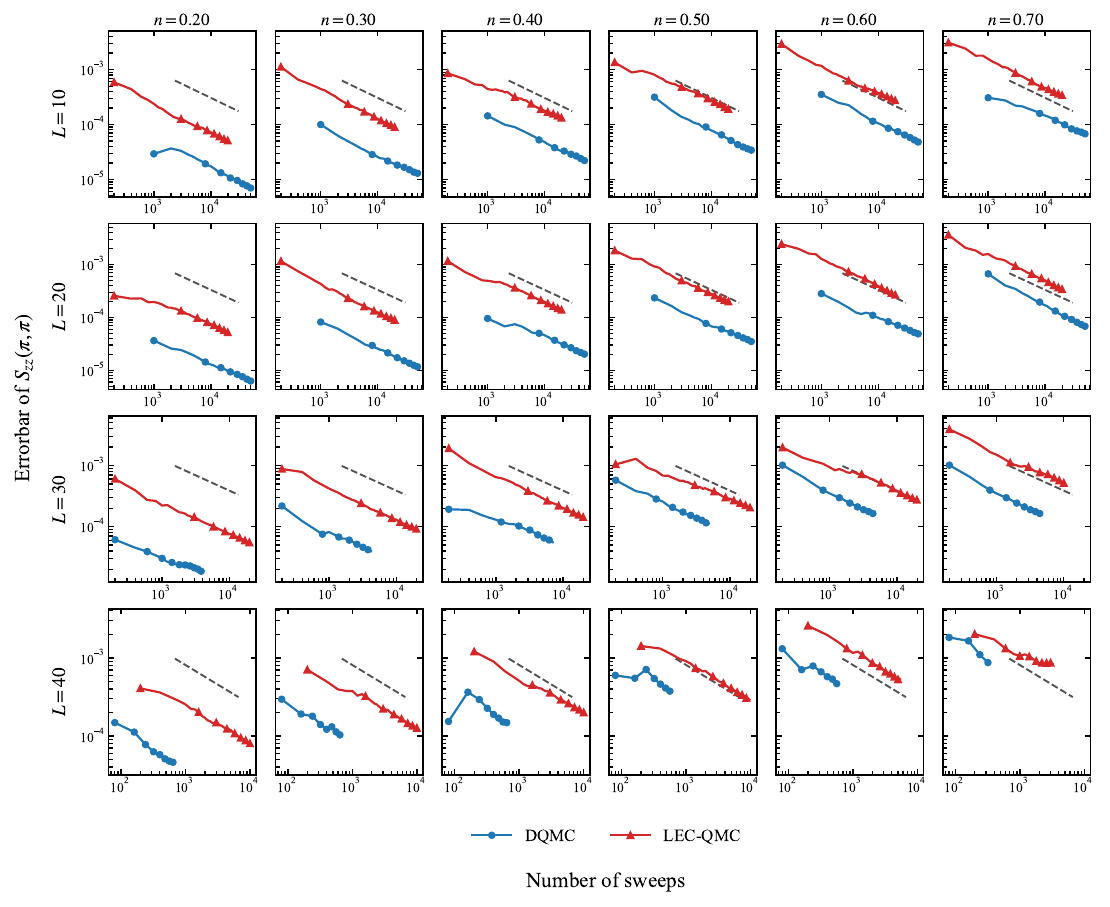}
    \caption{
    Statistical error bar of the antiferromagnetic spin structure factor $S^{zz}(\pi,\pi)$ as a function of the number of Monte Carlo sweeps, shown on a log-log scale. The parameters are the same as in Fig.~\ref{fig:energy_error_sweeps}. The gray dashed line indicates the ideal $N_{\rm sweep}^{-1/2}$ scaling expected for statistically independent samples. The per-sweep behavior is qualitatively similar to the energy benchmark.
    }
    \label{fig:szsz_error_sweeps}
\end{figure}

\begin{figure}[t]
    \centering
    \includegraphics[width=0.95\linewidth]{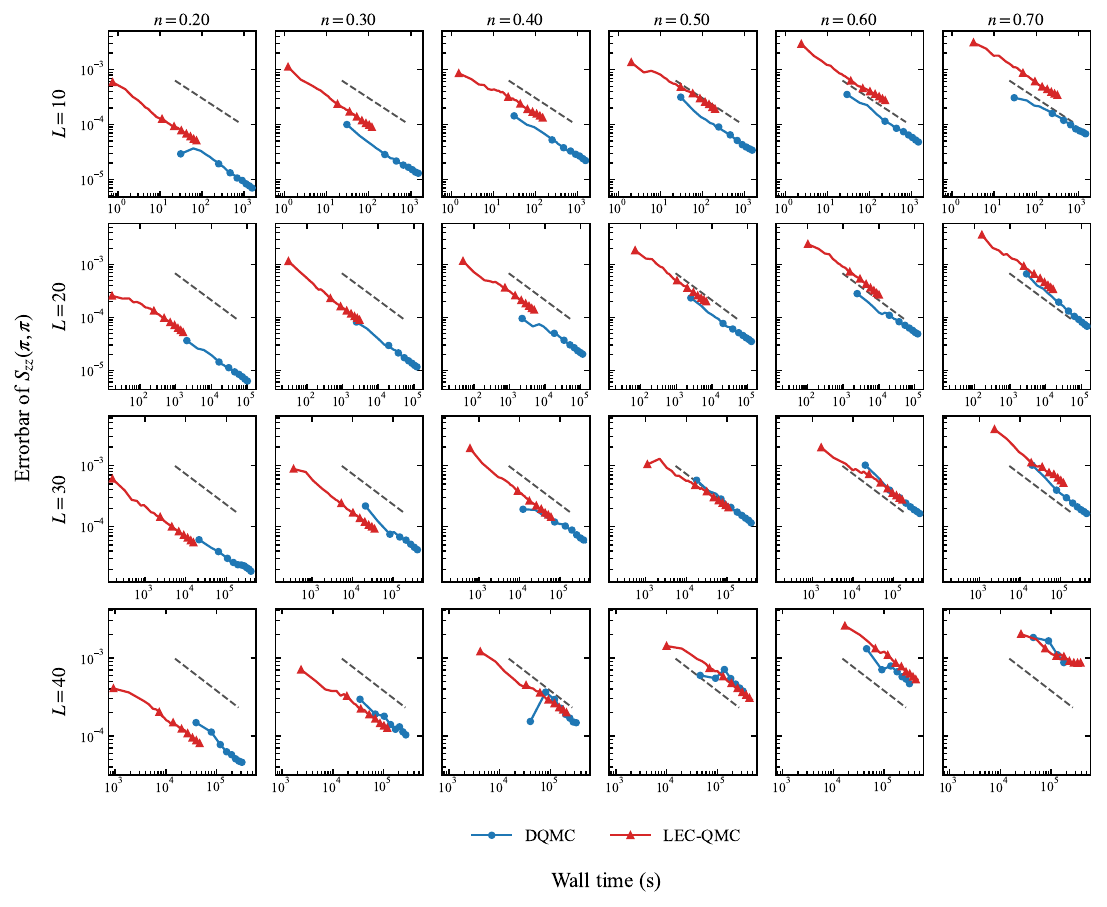}
    \caption{
    Statistical error bar of the antiferromagnetic spin structure factor $S^{zz}(\pi,\pi)$ as a function of wall-clock time, shown on a log-log scale. The gray dashed line indicates the ideal $t_{\rm wall}^{-1/2}$ scaling expected when the effective number of independent samples grows linearly with wall-clock time. Once the sweep cost is included, LEC-QMC becomes more efficient in the dilute and large-system regimes. The largest-size data have larger uncertainty because obtaining comparable DQMC statistics is computationally expensive.
    }
    \label{fig:szsz_error_time}
\end{figure}

Figs.~\ref{fig:szsz_error_sweeps} and~\ref{fig:szsz_error_time} show the error bar of $S^{zz}(\pi,\pi)$ as functions of the number of sweeps and wall-clock time, respectively. The overall trend is similar to that observed for the energy. At fixed sweep number, DQMC can have smaller error bars, particularly at higher filling. When the comparison is made at fixed wall-clock time, however, LEC-QMC becomes more efficient in the dilute and large-system regimes because of its much lower sweep cost.

Together, the energy and spin-correlation benchmarks show that LEC-QMC does not merely gain speed at the level of individual updates; its scaling advantage also translates into practical statistical efficiency for physical observables, especially in the regime $N_e\ll N$ where standard DQMC remains governed by the full system size.

\section{Construction of the flat band model}

In this section, we detail the construction of the flat-band model via the bipartite crystalline lattice (BCL) formalism and its chiral limit reduction proposed in Ref.~\cite{cualuguaru2022general}.

\subsection{Analysis of the general BCL Hamiltonian}
We consider a system with three bands, defined on a bipartite lattice structure. The basis is chosen such that the first two components correspond to the majority sublattice (site $A_1$ and site $A_2$), and the third component corresponds to the minority sublattice (site $B$). The Hamiltonian in momentum space is given by:

\begin{equation}
H(\vec{k})=\left(\begin{array}{ccc}
a & 0 & S_{1}(\vec{k})\\
0 & a & S_{2}(\vec{k})\\
S_{1}^{*}(\vec{k}) & S_{2}^{*}(\vec{k}) & B(\vec{k})
\end{array}\right)    
\end{equation}
where $a$ and $B(\vec{k})$ represent the onsite potentials (or intra-sublattice terms) for the two sublattices, and $S_{1,2}(\vec{k})$ describe the hopping between the sublattices.

By solving the characteristic equation, we obtain the three eigenvalues:

\begin{gather}
    \lambda_1 = a, \\
    \lambda_{2,3} = \frac{1}{2} \left[ a+ B(\vec{k}) \mp \sqrt{(a-B(\vec{k}))^2 + 4 (|S_1(\vec{k})|^2 + |S_2(\vec{k})|^2)} \right].
\end{gather}
The band $\lambda_1 = a$ is perfectly flat. For our purposes, we require this flat band to be the strictly lowest energy band. However, the general BCL construction described above cannot directly satisfy this condition with non-zero diagonal terms. As extensively discussed in Ref.~\cite{cualuguaru2022general}, to realize the flat band as the lowest band, one needs to consider a specific limit or ``integrate out'' certain degrees of freedom to obtain an effective low-energy model.

\subsection{The chiral limit and effective Hamiltonian}

To realize the flat band as the ground state, we consider the chiral limit by setting $a=0$ and $B(\vec{k}) = 0$. The Hamiltonian simplifies to a purely off-diagonal block form:

\begin{equation}
H_{\text{chiral}}(\vec{k}) = \begin{pmatrix} 
0 & 0 & S_1(\vec{k}) \\ 
0 & 0 & S_2(\vec{k}) \\ 
S_1^*(\vec{k}) & S_2^*(\vec{k}) & 0 
\end{pmatrix} = \begin{pmatrix} 
\mathbf{0}_{2\times2} & \mathbf{S}(\vec{k}) \\ 
\mathbf{S}^\dagger(\vec{k}) & 0 
\end{pmatrix},
\end{equation}
where $\mathbf{S}(\vec{k}) = [S_1(\vec{k}), S_2(\vec{k})]^T$ is a column vector.

To obtain the low-energy effective physics, we ``integrate out'' the minority sublattice (the third row/column). Mathematically, this is equivalent to considering the squared Hamiltonian $H^2$ or the projection onto the majority sublattice. The effective Hamiltonian for the two sites
in the majority sublattice corresponds to the $2 \times 2$ block $\mathbf{S}(\vec{k})\mathbf{S}^{\dagger}(\vec{k})$:

\begin{equation}
H_{\text{eff}}(\vec{k}) = \mathbf{S}(\vec{k}) \mathbf{S}^\dagger(\vec{k}) = \begin{pmatrix} 
|S_1(\vec{k})|^2 & S_1(\vec{k})S_2^*(\vec{k}) \\ 
S_2(\vec{k})S_1^*(\vec{k}) & |S_2(\vec{k})|^2 
\end{pmatrix}.
\end{equation}
The eigenvalues of this effective Hamiltonian are straightforwardly:
\begin{equation}
    \lambda_1 = 0, \quad \lambda_2 = |S_1(\vec{k})|^2 + |S_2(\vec{k})|^2.
\end{equation}
Here, $\lambda_1 = 0$ corresponds to the perfectly flat ground state, and $\lambda_2$ forms the dispersive excited band. This effective model $H_{\text{eff}}$ successfully isolates the flat band at zero energy.

We implement this scheme in a one-dimensional lattice model using the coupling functions:
\begin{equation}
    S_1(k) = t_1 + t_2 e^{-ik}, \quad S_2(k) = t_3 + t_4 e^{-ik},
\end{equation}
where $t_1$, $t_2$, $t_3$ and $t_4$ are all taken to be real for simplicity. In our simulations, we set $t_1 = t_4 = -0.2$ and $t_2 = t_3 = 1.0$, and a real-space sketch of the model is shown in Fig.~3 of the main text.

\section{Proof of Ferromagnetic Ground State}
In this section, we provide a rigorous proof that the ground state of the model illustrated in Fig.~3 of the main text is ferromagnetic. Following the theoretical framework established in Ref.~\cite{mielke1993ferromagnetism, tasaki1998nagaoka}, our derivation proceeds in two steps: we first construct the localized Wannier functions, and then apply the completeness and connectivity conditions to demonstrate the uniqueness of the ferromagnetic ground state.

\subsection{Exact construction of compact localized states}
To derive the explicit form of the Wannier functions in real space, we first solve for the zero-energy eigenstate in momentum space. The condition for a zero mode, $H_{\text{eff}}|\psi_{k}\rangle = 0$, implies $\mathbf{S}^{\dagger}(\vec{k})|\psi_{k}\rangle = 0$. Therefore, the components of $|\psi_k\rangle$ in plane wave basis ($|A_{1,k}\rangle,\ |A_{2,k}\rangle$) can have the following form up to a normalization factor,
\begin{equation}
    \psi_{A_1}(k) = S_{2}^{*}(k), \ \ \ \psi_{A_2}(k) = -S_{1}^{*}(k)
    \label{eq:zero_mode}
\end{equation}
To obtain the real-space Wannier function $|W_{R}\rangle$ centered at unit cell $R$, we perform the inverse Fourier transform:
\begin{equation}
    |W_{R}\rangle = \frac{1}{\sqrt{N}} \sum_{k} e^{-ikR} |\psi_{k}\rangle
    \label{eq:wannier}
\end{equation}
Substituting Eq.~(\ref{eq:zero_mode}) into Eq.~(\ref{eq:wannier}) and utilizing the relation between site basis and plane wave basis,
\begin{align}
\frac{1}{N}\sum_k e^{-ikR} |A_{1,k}\rangle & = |A_{1,R}\rangle, \\
\frac{1}{N}\sum_k e^{-ikR} |A_{2,k}\rangle & = |A_{2,R}\rangle,
\end{align} 
we obtain the site-resolved amplitudes:
\begin{equation}
|W_{R}\rangle=t_{3}|A_{1,R}\rangle+t_{4}|A_{1,R-1}\rangle-t_{1}|A_{2,R}\rangle-t_{2}|A_{2,R-1}\rangle
\end{equation}
Thus, the effective flat band Wannier function is a compact localized state confined strictly to two adjacent unit cells in the effective lattice, as depicted by the shaded region in Fig.~3. It is straightforward to verify that for this localized state, the total hopping amplitude to any outside site vanishes.

\subsection{Proof of ferromagnetism via Mielke-Tasaki mechanism}
We now address the magnetic ground state of the system when the flat band is half-filled (filling factor $\nu=1$, corresponding to $N_e = L$ electrons in a system of $L$ unit cells) in the presence of a repulsive Hubbard interaction $U>0$.

The stability of ferromagnetism in flat-band systems is governed by the connectivity of the local basis states, as established by Mielke and Tasaki. The proof relies on two conditions satisfied by our constructed basis $\{|W_{R}\rangle\}$:

\begin{enumerate}
    \item \textbf{Completeness:} The set of Wannier functions $\{|W_{R}\rangle\}$ is linearly independent and spans the entire flat band manifold.
    
    \item \textbf{Connectivity:} A crucial requirement for ferromagnetism is that the Wannier functions must overlap in real space to mediate exchange interactions.
    \begin{itemize}
        \item Consider the Wannier function $|W_{R}\rangle$, which has support on cells $\{R, R-1\}$.
        \item The adjacent function $|W_{R+1}\rangle$ has support on cells $\{R+1, R\}$.
        \item Since both functions share support on the sites of unit cell $R$ (specifically sites $A_{1,R}$ and $A_{2,R}$), their overlap integral is:
        \begin{equation}
        \langle W_{R} | W_{R+1} \rangle = t_1 t_2 + t_3 t_4
        \end{equation}
    \end{itemize}
    As long as we choose parameters $t_1 t_2 + t_3 t_4 \ne 0$, their overlap is non-vanishing.
\end{enumerate}

This non-vanishing overlap implies that the ground state manifold is connected. According to Mielke and Tasaki's theorems~\cite{mielke1993ferromagnetism} on positive semi-definite Hamiltonians, when the single-particle ground states are degenerate and satisfy this connectivity condition, the many-body ground state at half-filling is the unique ferromagnetic state with maximal total spin $S_{\text{tot}} = N_{e}/2$, apart from the trivial $(2S_{\text{tot}}+1)$-fold spin rotation degeneracy.

Any state with lower total spin would require constructing spatial domain walls or occupying higher energy bands to satisfy the Pauli exclusion principle, thereby incurring an energy penalty. Therefore, the system exhibits robust ferromagnetism driven by the geometric frustration and quantum interference inherent in the flat band.

Due to the $\text{SU}(2)$ spin-rotation symmetry of the Hamiltonian, the ferromagnetic ground state possesses a $(2S_{\text{tot}}+1)$-fold degeneracy. However, in our canonical ensemble simulations, we restrict the particle numbers such that $N_{\uparrow} = N_{\downarrow} = N_e/ 2$. This constraint selects the $S^z_{\text{tot}} = 0$ sector, effectively singling out a specific state within the ferromagnetic multiplet. Consequently, the observed ferromagnetic order manifests within the XY-plane, as shown in the main text.

\section{EXACT CALCULATION OF SUBLATTICE SPIN FLUCTUATIONS}

In this section, we provide a rigorous derivation of the squared sublattice magnetization $\langle (S_{A}^z)^2 \rangle$ for the ferromagnetic ground state in the spin sector $S_{\text{tot}}^z = 0$. This analytic result rationalizes the convergence value observed in Fig.~4(b).

\subsection{Formalism}

We consider a system of $N_e$ electrons filling a completely flat band. The many-body ground state $|\Psi_0\rangle$ exhibits perfect ferromagnetism. In the sector with zero total $z$-component spin $S_{\text{tot}}^z = 0$, the wavefunction factorizes into a spin part and a spatial part: the spin component corresponds to the Dicke state $|S=N_e/2, M=0\rangle$, while the spatial component is a Slater determinant formed by occupying all Bloch states $|\psi_{k}\rangle$ in the flat band.

The total spin-$z$ operator on sublattice $A_1$ is defined as:
\begin{equation}
    S_{A_1}^z = \sum_{i=1}^{N_e} \hat{s}_{i}^z \hat{\mathcal{P}}_{A_1}(i),
\end{equation}
where $\hat{s}_{i}^z$ is the spin operator for the $i$-th electron, and $\hat{\mathcal{P}}_{A_1}(i)$ is the projection operator onto sublattice $A_1$ for the $i$-th electron.

We aim to calculate the fluctuation $\langle (S_{A_1}^z)^2 \rangle = \langle \Psi_0 | (S_{A_1}^z)^2 | \Psi_0 \rangle$. Expanding the square yields single-particle and two-particle terms:
\begin{align}
    (S_{A_1}^z)^2 &= \sum_{i,j} \hat{s}_{i}^z \hat{s}_{j}^z \hat{\mathcal{P}}_{A_1}(i) \hat{\mathcal{P}}_{A_1}(j) \nonumber \\
    &= \sum_{i} (\hat{s}_{i}^z)^2 \hat{\mathcal{P}}_{A_1}(i) + \sum_{i \neq j} \hat{s}_{i}^z \hat{s}_{j}^z \hat{\mathcal{P}}_{A_1}(i) \hat{\mathcal{P}}_{A_1}(j).
\end{align}
Due to the decoupled nature of spin and spatial degrees of freedom in the flat band ferromagnet, we can evaluate the expectation values of the spin and orbital operators separately.

\subsection{Spin correlations}

For the Dicke state $|S=N_e/2, M=0\rangle$, the onsite spin fluctuation is $\langle (\hat{s}_{i}^z)^2 \rangle = \frac{1}{4}$. The total spin fluctuation vanishes, $\langle (\hat{S}_{\text{tot}}^z)^2 \rangle = \sum_{i,j} \langle \hat{s}_{i}^z \hat{s}_{j}^z \rangle = 0$, which implies a sum rule for the pair correlations:
\begin{equation}
    \sum_{i \neq j} \langle \hat{s}_{i}^z \hat{s}_{j}^z \rangle = - \sum_{i} \langle (\hat{s}_{i}^z)^2 \rangle = -\frac{N_e}{4}.
\end{equation}
By symmetry among identical fermions, the correlation between any distinct pair $i \neq j$ is uniform:
\begin{equation}
    \langle \hat{s}_{i}^z \hat{s}_{j}^z \rangle = -\frac{1}{4(N_e-1)}.
\end{equation}

\subsection{Spatial correlations in momentum space}

The calculation of spatial projections is most conveniently performed in reciprocal space, where the Bloch basis is orthogonal. Let $p(k)$ denote the probability weight of the flat band eigenstate on sublattice $A_1$ at momentum $k$:
\begin{equation}
    p(k) = \langle \psi_k | \hat{\mathcal{P}}_{A_1} | \psi_k \rangle = \frac{|S_2(k)|^2}{|S_1(k)|^2 + |S_2(k)|^2},
\end{equation}
where $S_1(k)$ and $S_2(k)$ are the structure factors defined previously.

We define the moments of the sublattice weight as $P_n = \sum_k [p(k)]^n$. The expectation value of the single-particle projector is simply:
\begin{equation}
    \sum_{i} \langle \hat{\mathcal{P}}_{A_1}(i) \rangle = \sum_k p(k) = P_1.
\end{equation}
For the two-particle term, since the spatial wavefunction is a single Slater determinant of orthogonal Bloch states, the joint probability factorizes for distinct momenta:
\begin{align}
    \sum_{i \neq j} \langle \hat{\mathcal{P}}_{A_1}(i) \hat{\mathcal{P}}_{A_1}(j) \rangle &= \sum_{k \neq q} p(k) p(q) \nonumber \\
    &= \left( \sum_k p(k) \right)^2 - \sum_k \left[p(k)\right]^2 \nonumber \\
    &= P_1^2 - P_2.
\end{align}
The term $P_2$ captures the momentum-space texture of the wavefunction arising from the interference between the hopping amplitudes.

\subsection{Result}

Substituting the spin and spatial components back into the expansion, we obtain the general formula:
\begin{align}
    \langle (S_{A_1}^z)^2 \rangle &= \frac{1}{4} P_1 + \left( -\frac{1}{4(N_e-1)} \right) (P_1^2 - P_2) \nonumber \\
    &= \frac{1}{4(N_e-1)} \left[ (N_e-1)P_1 - P_1^2 + P_2 \right].
    \label{eq:SzA}
\end{align}

For the specific parameters used in our simulations ($t_1 = t_4 = t'$ and $t_2 = t_3 = t$), the Hamiltonian possesses a symmetry that enforces equal sublattice weights, $p(k) = 1/2$, independent of $k$. Consequently, the moments become $P_1 = N_e/2$ and $P_2 = N_e/4$. Substituting these into the general formula yields:
\begin{gather}
    \langle (S_{A_1}^z)^2 \rangle = \frac{N_e}{16}, \\
    \langle S^{zz}(q=0)\rangle_{11} \equiv \frac{1}{L} \langle (S_{A_1}^z)^2 \rangle = \frac{1}{16},
\end{gather}
where we have used the half-filling condition $N_e = L$. This analytical value of $1/16$ is in perfect agreement with the asymptotic behavior observed in Fig.~4(b).

We further validated Eq.~(\ref{eq:SzA}) using Exact Diagonalization (ED) for arbitrary sets of hopping parameters $t_{1,2,3,4}$ where $p(k)$ is not constant. The numerical results show perfect agreement with our analytical derivation across all parameter regimes.

\section{A SIMPLE CONSTRAINED PATH FRAMEWORK FOR LEC-QMC}

In this section, we provide a concise formulation of the constrained path (CP) realization within the LEC-QMC framework. The introduction of the constrained path approximation extends the applicability of LEC-QMC to arbitrary parameter regimes by bypassing the sign problem, while fully preserving its linear scaling complexity. The framework we propose combines the CP approximations in PQMC and finite-temperature DQMC (FT-DQMC)~\cite{zhangConstrainedPathMonte1997, he2019finite}.

\subsection{Two-Stage Sampling Scheme}

We consider the Hubbard model as an example,
\begin{equation}
  H = -t \sum_{\langle i,j \rangle, \sigma} \left( c_{i\sigma}^\dagger c_{j\sigma} +\text{h.c.} \right) + U \sum_i n_{i\uparrow} n_{i\downarrow}.
\end{equation}
The canonical partition function for $N_e$ fermions is written as:
\begin{equation}
Z = \text{Tr}\left( e^{-\beta \hat{H}} \right) = \sum_{\eta} \langle \eta | e^{-\beta \hat{H}} | \eta \rangle.
\end{equation}
To circumvent the sign problem without performing an explicit summation over all Fock states $\eta$, we introduce a reweighting procedure based on the trial Hamiltonian $\hat{H}_0$. Defining the trial weight as $W_T(\eta) = \langle \eta | e^{-\beta \hat{H}_0} | \eta \rangle$, we rewrite the partition function as:
\begin{equation}
    Z = \sum_{\eta} W_T(\eta) \times \left[ \frac{\langle \eta | e^{-\beta \hat{H}} | \eta \rangle}{\langle \eta | e^{-\beta \hat{H}_0} | \eta \rangle} \right].
\end{equation}
By applying the Hubbard-Stratonovich (HS) transformation to the interaction term, the quantity in the brackets can be expressed as a path integral over the auxiliary fields $\bm{s}$:
\begin{equation}
    \mathcal{R}(\eta) = \frac{\int \mathcal{D}\bm{s} \, P(\bm{s}) \, \langle \eta | \hat{U}(\bm{s}) | \eta \rangle}{\langle \eta | e^{-\beta \hat{H}_0} | \eta \rangle},
\end{equation}
where $P(\bm{s})$ is a scalar factor which comes from HS transformation and depends on auxiliary fields, and $\hat{U}(\bm{s}) = \prod_{l=1}^{L_\tau} e^{-\Delta\tau \hat{H}_0} e^{-\Delta\tau \hat{H}_I(\bm{s}_l)}$ is the time-discretized propagator.

This decomposition justifies a two-level Markov Chain Monte Carlo approach:

\noindent \textbf{1. Outer Loop (Fock State Sampling):} Sample Fock states $\eta$ according to the trial distribution $\mathcal{P}(\eta) \propto W_T(\eta) = \langle \eta | e^{-\beta \hat{H}_0} | \eta \rangle$. This is efficiently performed using the standard LEC-QMC algorithm driven by the sign-problem-free $\hat{H}_0$.

\noindent \textbf{2. Inner Loop (Auxiliary Field Sampling):} For a fixed Fock state $\eta$, sample the auxiliary fields $\bm{s}$ according to the weight
\begin{equation}
 \frac{P(\bm{s}) \, \langle \eta | \hat{U}(\bm{s}) | \eta \rangle}{\langle \eta | e^{-\beta \hat{H}_0} | \eta \rangle}.
\end{equation}
The constrained path approximation is applied within this loop to control the phase of the walker weights.

\subsection{Random Walkers and Linear Scaling Updates}

In the inner loop, we evaluate the integral for $\mathcal{R}(\eta)$ using a branching random walk. The ratio $\mathcal{R}(\eta)$ is factorized into a product of time-step updates:
\begin{gather}
    \mathcal{P}_l^T = \langle \eta| e^{-(\beta-\tau_l) \hat{H}_0 } \hat{U}(\bm{s}_{\tau_l \dots \tau_1}) | \eta \rangle, \\
    \mathcal{R}(\eta) \approx \frac{\mathcal{P}^T_{L_\tau}}{\mathcal{P}^T_{L_\tau-1}} \frac{\mathcal{P}^T_{L_\tau-1}}{\mathcal{P}^T_{L_\tau-2}} \cdots \frac{\mathcal{P}^T_1}{\mathcal{P}^T_{0}}.
\end{gather}
We define a walker $|\phi_l\rangle$ at time slice $\tau_l$ as the state evolved from the initial $|\eta\rangle$ under the interacting propagator. The importance sampling is guided by the trial overlap $O_T(\tau_l)$, defined as the projection of the walker onto the backward-propagated trial state $\langle \psi(\tau_l) | = \langle \eta | e^{-(\beta - \tau_l)\hat{H}_0}$:
\begin{equation}
    O_T(\tau_{l}) = \langle \psi(\tau_l) | \phi_l \rangle = \det[ \Psi^{\dagger}(\tau_l)\Phi_l ],
\end{equation}
where $\Phi_l$ is the $N \times N_{e}$ matrix representation of the walker, and $\Psi^{\dagger}(\tau_l)$ represents the backward trial state.

To mitigate the sign problem, we enforce the phaseless approximation, requiring the walker to maintain a positive overlap with the guiding trial wavefunction. At each propagation step $l \rightarrow l+1$, the walker's weight is updated by the real part of the overlap ratio:
\begin{equation}
w_{l+1} = w_l \times \max\left(0, \text{Re}\left[ \frac{O_T(\tau_{l+1})}{O_T(\tau_l)} \right] \right).
\end{equation}
Walkers with non-positive effective weights are discarded. This constraint effectively truncates paths that cross the nodal surface defined by $\hat{H}_0$.

To preserve the $O(N)$ linear scaling of LEC-QMC while minimizing variance, we perform auxiliary field updates site-by-site. This sequential strategy is essential because the optimal sampling probability for site $i+1$ depends on the walker's configuration updated at site $i$.

The update of the auxiliary field $s_i$ at site $i$ involves applying a local operator $e^{-\Delta\tau \hat{H}_I(s_i)}$, which induces a rank-1 update to the walker matrix $\Phi$:
\begin{gather}
\Phi' = (1 + \Delta_i) \Phi, \\
r = \frac{\det(\Psi^\dagger (1+{\Delta}_i) \Phi)}{\det(\Psi^\dagger \Phi)}.
\end{gather}
For numerical stability, we perform QR decompositions on both the backward trial state matrix $\Psi^\dagger = D_L U_L^\dagger$ and the walker matrix $\Phi = U_R D_R$, where $U_{L/R}$ are unitary matrices and $D_{L/R}$ are upper-triangular. The large scales in $D_{L/R}$ cancel out in the ratio, leaving only the well-conditioned unitary components:
\begin{equation}
r = \frac{\det( D_L U_L^\dagger (1+{\Delta}_i) U_R D_R)}{\det(D_L U_L^\dagger U_R D_R)} = \frac{\det(U_L^\dagger (1+{\Delta}_i) U_R)}{\det(U_L^\dagger U_R)}.
\end{equation}
This formulation allows the ratio $r$ and the matrix updates to be computed efficiently using the Sherman-Morrison formula. Consequently, the computational cost per sweep scales linearly with system size, consistent with the standard LEC-QMC.

We note that while advanced techniques such as force bias or heat-bath importance sampling are compatible with this framework, we present the simplest realization here to demonstrate the core methodology.

\newpage
\bibliography{LEC-QMC.bib}